\documentclass{emulateapj}
\usepackage{apjfonts,graphics}
\usepackage{amssymb}
\usepackage{alltt}
\usepackage{epsfig}
\usepackage{subfigure}
\usepackage{lscape}
\usepackage{longtable}
\usepackage{textcomp}
\usepackage{graphicx}
\usepackage{natbib}

\newcommand{\be}{\begin{equation}}
\newcommand{\ee}{\end{equation}}
\newcommand{\ba}{\begin{eqnarray}}
\newcommand{\ea}{\end{eqnarray}}

\begin{document}

\title{Using {\it Weak Lensing Dilution} to Improve Measurements of
  the Luminous and Dark Matter in A1689}

\author{Elinor Medezinski\altaffilmark{1}, Tom
  Broadhurst\altaffilmark{1}, Keiichi Umetsu\altaffilmark{2}, Dan Coe
  \altaffilmark{3,4}, Narciso Ben\'\i tez\altaffilmark{4}, Holland
  Ford\altaffilmark{3}, Yoel Rephaeli \altaffilmark{1}, Nobuo
  Arimoto\altaffilmark{5}, Xu Kong\altaffilmark{6} }

\altaffiltext{1}{School of Physics and Astronomy, Tel Aviv University,
  Israel} 
\altaffiltext{2}{Institute of Astronomy and Astrophysics,
  Academia Sinica, P.~O. Box 23-141, Taipei 106, Taiwan, Repucblic of China}
\altaffiltext{3}{Physics and Astronomy Dept, Johns Hopkins
  University, Baltimore Maryland, USA} 
\altaffiltext{4}{Instituto de
  Astrof\'\i sica de Andaluc\'\i a (CSIC), Granada, Spain}
\altaffiltext{5}{National Astronomical Observatory of Japan, Mitaka
  181-8588, Japan} 
\altaffiltext{6}{Center for Astrophysics, University of
  Science and Technology of China, Hefei, 230026, Anhui, PR China}

\begin{abstract}  
  The E/SO sequence of a cluster defines a boundary redward of which a
  reliable weak lensing signal can be obtained from background
  galaxies, uncontaminated by cluster members. For bluer colors, both
  background and cluster galaxies are present, reducing the average
  distortion signal by the proportion of unlensed cluster members. In
  deep Subaru and HST/ACS images of A1689, we show that the tangential
  distortion of galaxies with bluer colors falls rapidly toward the
  cluster center relative to the background reference level provided
  by the red background. We use this dilution effect to derive the
  cluster light profile and luminosity function to large radius, with
  the advantage that no subtraction of far-field background counts is
  required. The light profile of A1689 is found to decline steadily to
  the limit of the data, $r<2~h^{-1}$Mpc, with a constant slope,
  $d\log(L)/d\log(r)=-1.12\pm0.06$, unlike the lensing mass profile
  which steepens continuously with radius, so that $M/L$ peaks at an
  intermediate radius, $\simeq 100~h^{-1}$ kpc. A flatter behaviour is
  found for the more physically meaningful ratio of dark-matter to
  stellar-matter, when account is made of the color-mass relation of
  cluster members. We derive a cluster luminosity function with a flat
  faint-end slope of $\alpha=-1.05\pm0.07$, nearly independent of
  radius and with no faint upturn to $M_{i'}<{-12}$. We establish that
  the very bluest objects are negligibly contaminated by the cluster
  ($[V-i']_{AB}<0.2$), because their distortion profile rises
  continuously towards the center following the red background
  ($[V-i']_{AB}>1.2$), but is offset higher by $\simeq 20\%$. This
  larger amplitude is consistent with the greater estimated depth of
  the faint blue galaxies $\langle z\rangle\sim2$ compared to $\langle
  z\rangle\sim0.85$ for the red background, a purely geometric effect.
  With a larger sample of background galaxies behind several clusters
  we may use this geometric effect to constrain the cosmological
  parameters in a model independent way, by comparing the weak lensing
  strength over a wide range of photometrically derived redshifts.
  Finally, we improve upon our earlier modeling of the mass profile by
  combining both the red and blue background populations, and very
  clearly exclude low concentration profiles predicted for massive CDM
  dominated haloes.
\end{abstract}                   


\section{Introduction}\label{Intro}

The influence of ``Dark matter'' is strikingly evident in the centers
of massive galaxy clusters, where large velocity dispersions are
measured and giant arcs are often visible.  Cluster masses may be
estimated by several means, leading to exceptionally high central
mass-to-light ratios, $M/L_R \sim 100-300h(M/L_R)_{\odot}$
\citep{2001ApJ...552..427C}, far exceeding both the mass of stars
comprising the light of the cluster galaxies and the mass of plasma
derived from X-ray emission and the SZ effect.  Reasonable consistency
is claimed between dynamical, hydrodynamical and lensing-based
estimates of cluster masses, supporting the conventional understanding
of gravity.  However, the high mass-to-light ratio requires that an
unconventional non-baryonic dark material dominates the mass of
clusters, whose origin remains very unclear.

In detail, discrepancies are often reported between masses derived
from strong lensing and X-ray measurements, with the claimed X-ray
masses often lower in the centers of clusters. High resolution X-ray
emission and temperature maps reveal that the majority of local
clusters undergo repeated merging with sub-clumps, and obvious shock
fronts are seen in some cases
\citep{2002ApJ...567L..27M,2004ApJ...608..179R}. The double cluster
1E~0657--56 (a.k.a., the ``bullet'' cluster, $z=0.296$) is the most
extreme example studied, where the associated gas forms a flattened
luminous shock-heated structure lying between the two large distinct
clusters, clearly indicating these two bodies collided recently at a
high relative velocity \citep{2004ApJ...606..819M}, with the gas
remaining in between while the cluster galaxies have passed through
each other relatively collisionlessly. Very interestingly, the weak
lensing signal follows the double structure of the clusters, rather
than the gas in between, cleanly demonstrating that the bulk of the
matter is collisionless and dark \citep{2004ApJ...604..596C}. This
system favors the standard cold dark matter (CDM) scenario, places a
restrictive limit on the interaction cross-section of any fermionic
dark matter, and disfavors a class of alternative gravity theories in
which only baryons are present \citep{2004ApJ...604..596C}. Smaller
but significant discrepancies of the same sort are also claimed for
other interacting clusters
\citep{2002ApJ...580L..17N,2005ApJ...618...46J}.

Many clusters show no apparent signs of significant ongoing
interaction; these have centrally symmetric X-ray emission and little
obvious substructure \citep{1998MNRAS.296..392A}, and for some of
these the lensing and X-ray (or dynamical) derived masses are claimed
to agree, as would be expected for relaxed systems
\citep[e.g.,][]{2002ApJ...572...66A,2003AJ....126.2152R,2005ApJ...628L..97D}.
More generally, since the dyanmics of dark matter and most cluster
galaxies is essentially collisionless, we would expect them to have
similar radial profiles.  Biasing inherent in hierarchical growth may
significantly modify this similarity \citep{1997MNRAS.286..795K}, and
dynamical friction is expected to concentrate the relatively more
massive galaxies in the core. These together with tidal effects may
help to account for the unique properties of cD galaxies. Hence
comparisons of the mass profile with the light profile of the cluster
galaxies are expected to provide an additional insight into the
formation of clusters and the nature of dark matter.

Recent improvements in the quality of data useful for weak and strong
lensing studies now allow the construction of much more definitive
mass profiles that are sufficiently precise to test the predictions of
popular models, relatively free of major assumptions.  The inner mass
profiles of several clusters have been constrained in some detail via
lensing, using multiply lensed background galaxies
\citep{1996ApJ...471..643K,1997ApJ...491..477H,2002ApJ...574L.129S,
  2003A&A...403...11G,2004ApJ...604...88S,2005ApJ...621...53B,
  2005ApJ...629L..73S}. The statistical effects of weak lensing have
been used to extend the mass profile to larger radii. The mass
profiles derived from these observations have been claimed to show
NFW-like behaviour \citep*{1997ApJ...490..493N}, with a continuous
flattening towards the center, but with higher density concentrations
than expected. This is particularly evident for A1689, for which we
have constructed the highest quality lensing based mass profile to
date, combining over 100 multiply-lensed images and weak lensing
effects of distortion and magnification from Subaru
\citep{2005ApJ...621...53B,2005ApJ...619L.143B}.

A lesson learned from this earlier work is the importance of carefully
selecting a background population to avoid contamination by the
lensing cluster. It is not enough to simply exclude a narrow band
containing the obvious E/SO sequence, following common practice,
because the lensing signal of the remainder is found to fall rapidly
towards the cluster center, in contrast to the uncontaminated
population of background galaxies lying redward of the cluster sequence
\citep{2005ApJ...619L.143B}. In the well studied case of A1689, there
has been a long standing discrepancy between the strong and weak
lensing effects, with the weak lensing signal underpredicting the 
observed Einstein radius by a factor of $\sim 2.5$ 
\citep{2001A&A...379..384C,2005A&A...434..433B}, based only on a 
minimal rejection of obvious cluster members using one or two-band 
photometry.

In our recent weak lensing analysis of Subaru images of A1689 the
above behaviour was found when we rejected only the cluster sequence
in the same way as others. This resulted in a relatively shallow trend
of the weak lensing signal with radius and consequently an
underprediction of the Einstein radius accounting for the earlier
discrepancy \citep{2005ApJ...619L.143B} . If, however, one selects
only objects redder than the cluster sequence for the lensing
analysis, then the weak tangential distortion continues to rise all
the way to the Einstein radius in very good agreement with the strong
lensing strength. This red population is naturally expected to
comprise only background galaxies, made redder by relatively large
k-corrections and with negligible contamination by cluster members,
since the bulk of the reddest cluster members are the early-type
galaxies defined by the cluster sequence. However, for galaxies with
colors bluer than the cluster sequence, cluster members will be
present along with background galaxies since the cluster population
extends to bluer colors of the later-type members, overlapping in
color with the blue background. The effect of the cluster members is
simply to reduce the strength of the weak lensing signal when averaged
over a statistical sample, in proportion to the fraction of cluster
members whose orientations are randomly distributed, therefore
diluting the lensing signal relative to the reference background level
derived from the red background population.

We can turn this dilution effect to our advantage and use it to derive
properties of the cluster population, in particular the radial light
profile, for comparison with the dark matter profile. Deriving a light
profile this way has advantages over the usual approach to defining
cluster membership.  The inherent fluctuations in the number counts of
the background population are a significant source of uncertainty in
the usual approach of subtracting the far-field level when defining
the cluster population
\citep[e.g.,][]{2001A&A...367...59P,2005MNRAS.360..727A,2005MNRAS.364.1147P}. This uncertainty is often cited as a
potential explanation for the substantial variation reported between
luminosity functions derived for different clusters, particularly in
the outskirts of clusters, where not only is the density of galaxies
lower, but their colors are bluer, thus harder to distinguish from
the background using photometry alone.

In \S~\ref{data} we describe the observations and photometry of the
Subaru images of A1689. In \S~\ref{dist_Sub} we describe the
distortion analysis applied to the Subaru data. 
The distortion analysis of ACS images of  A1689 is described in 
\S~\ref{dist_ACS}. 
In \S~\ref{Photo-z} we describe the photometric redshift analysis of the
Subaru and ACS images with reference to the \cite{2004AJ....127..180C}
sample of deep multi-color Subaru images. Our weak lensing dilution
analysis of the Subaru images is described in \S~\ref{dilution}. 
In \S~\ref{lum_sec} we go on to derive the cluster luminosity profile and 
color, and in \S~\ref{lum_func} the cluster luminosity function is 
deduced at several radial positions. In \S~\ref{ML_sec} we determine the 
$M/L$ profile, and in \S~\ref{MLconsist} we do a consistency check for 
the mass derived in this paper with previous estimations. 
Our conclusions are summarized in \S~\ref{discussion}. 

The concordance $\Lambda$CDM cosmology is adopted ($\Omega_M=0.3$, 
$\Omega_{\Lambda}=0.7$ but $h$ is left in units of $H_0/100$ km s$^{-1}$ 
Mpc$^{-1}$, for easier comparison with earlier work).

\section{Subaru Imaging reduction and Sample selection}  
\label{data}

We have retrieved Suprime-Cam imaging of A1689 in $V$ (1920s) and SDSS
$i'$ (2640s) from the Subaru archive, SMOKA.
\footnote{http://smoka.nao.ac.jp.}
Reduction software developed by \cite{2002AJ....123...66Y} is used for
flat-fielding, instrumental distortion correction, differential
refraction, PSF matching, sky subtraction, and stacking. The resulting
FWHM is 0\arcsec.82 in $V$ and 0\arcsec.88 in $i'$ with 0\arcsec.202
pix$^{-1}$, covering a field of $30'\times 25'$.

\begin{figure}[tb]
  \centering \includegraphics[scale=.50]{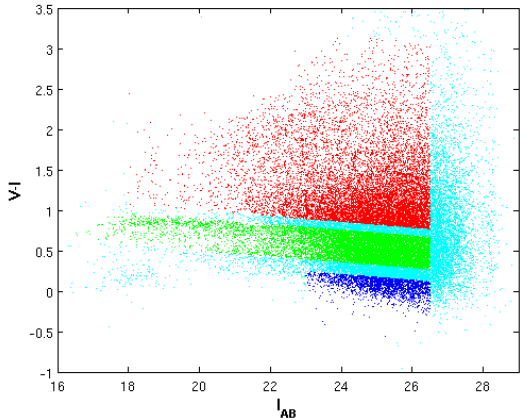}
  \caption{Color vs. magnitude diagram for A1689 cluster galaxies. The
    E/SO sequence is apparent at $(V-i')\sim0.8$ where there is an
    overdensity of bright galaxies. The red points represent the
    background sample of galaxies redder than the E/SO sequence. The
    blue points represent a background sample of the bluest objects in
    the field.  The green points cover a range of color chosen to
    include the cluster sequence and bluer cluster members, but in
    addition background galaxies are also present whose colors fall in
    this range.}  \label{colormag}
\end{figure}

\begin{table*}
\begin{center}
\caption{Sample Selection}\label{sample}
\begin{tabular}{lcccccc}
\hline\hline
Sample Name & mag limit & color limit & N & n [$h^2$ Mpc$^{-2}$]& $<z>$ & $<D>$ \\\hline
red & $18<i'<26.5$ & $0.2<(V-i')-(V-i')_{E/SO}<2.5$ & 11774 & 966.1 &
0.871$\pm$0.045 & 0.693$\pm$0.012 \\
green &  $16<i'<26.5$ & $-0.3<(V-i')-(V-i')_{E/SO}<0.1$ & 11963 &
981.6 & 1.429$\pm$0.093 & 0.728$\pm$0.015 \\
blue & $23<i'<26.5$ & $-1.2<(V-i')-(V-i')_{E/SO}<-0.45$ & 2459 &
201.8 & 2.012$\pm$0.124 & 0.830$\pm$0.011\\
\hline
\end{tabular}
\end{center}
\end{table*}

\begin{figure}[tb]
  \centering \includegraphics[scale=.50]{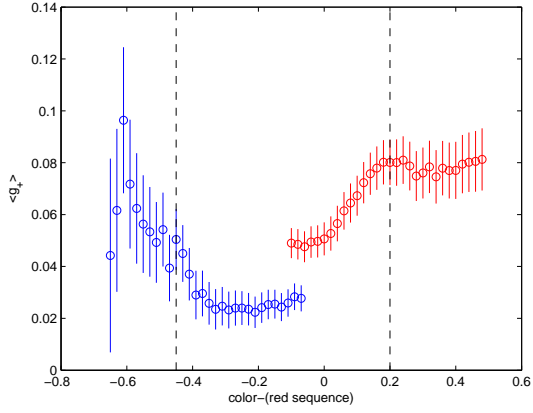}
  \caption{To establish the boundaries of the color distribution free
    of cluster members we calculate the mean tangential distortion
    averaged over the full radial extent of the cluster, done
    separately for the blue and red samples. On the right the red
    curve shows that $g_T$ drops rapidly when the bluer limit of the
    entire red sample is decreased below a color while lies $+0^m.2$
    magnitudes redward of the cluster sequence. This sharp decline
    marks the point at which the red sample encroaches on the E/SO
    sequence of the cluster. On the blue side the red limit of the
    blue sample is chosen to lie $-0^m.45$ blueward of the cluster
    sequence, marking the point at which significant contamination by
    the cluster acts to dilute the weak lensing signal from later-type
    bluer cluster members.}
  \label{limits}
\smallskip
\end{figure}

Photometry is based on a combined $V+i'$ image using SExtractor
\citep{1996A&AS..117..393B}. The limiting magnitudes are $V=26.5$ and
$i'=25.9$ for a $3\sigma$ detection within a $2$\arcsec aperture.  We
define three galaxy samples according to color and magnitude --
``red'', ``green'', and ``blue'' (see table~\ref{sample} for summary), and
for all our samples we define a limiting magnitude of $i'<26.5$, to
avoid incompleteness, as shown in Figure~\ref{colormag}. The red
galaxy sample consists of galaxies $0^m.2$ redder than the E/SO 
sequence of the cluster, which is accurately
defined by the linear relation $(V-i')_{E/SO}=-0.03525i'+1.505$, and
up to $2^m.5$ redder than this line to include the majority of the
background red population. Very red dropout galaxies may be detected
beyond this point. Indeed, one spectroscopically confirmed example at
$z=4.82$ has been detected behind this cluster
\citep{2002ApJ...568..558F}, and such cases are excluded by this upper 
limit, so that we do not need to make an uncertain correction for the 
level of their weak lensing signal which will be significantly larger 
than for the bulk of the background red galaxy population. As we will 
show in \S~\ref{Photo-z}, most of these red background galaxies are at 
a much lower mean redshift of $\langle z\rangle\sim0.85$. The red 
galaxy sample is redder than the cluster
sequence, made so by relatively large k-corrections, being largely
comprised of early to mid-type galaxies at moderate redshift (see
\S~\ref{Photo-z}). Cluster members are not expected to extend to these
colors in any significant numbers because the intrinsically reddest
class of cluster galaxies, E/SO galaxies, are defined by the cluster
sequence and lie comfortably blueward of the chosen sample limit (see
Fig.~\ref{colormag}), so that even large photometric errors will not
carry them into the red sample.  This can be demonstrated readily, as
shown in Figure~\ref{limits}, where we plot the mean lensing strength
as a function of color by moving the lower color limit progressively
blueward, finding a sharp drop in the lensing signal at our limit,
$(V-i')<(V-i')_{E/SO}+0.2$, when the cluster sequence starts to
contribute significantly, thereby reducing the mean lensing signal.

We define the blue galaxy sample as objects $0^m.45$ bluer than the
sequence line, with the magnitude limit in the interval $23<i'<26.5$,
so as to take only the very faint blue galaxies which - as we
establish below - are also negligibly contaminated by the cluster,
with a weak lensing signal which has the same radial dependence as the
red galaxy sample. We have explored the definition of the blue sample
when we realized that the bluest objects in the field have a
continuously rising weak lensing signal towards the cluster center
like the red galaxies, so that the ``contamination'' is minimal with
an insignificant effect on the quantities of interest for our
purposes. Figure~\ref{limits} shows that as the blue sample upper
color limit is advanced redwards, the integrated strength of the mean
weak lensing signal declines markedly within $0^m.45$ of the cluster
sequence, at $(V-i')>(V-i')_{E/SO}-0.45$. The reduction in signal is
more gradual than for the red population (both illustrated in
Fig.~\ref{limits}) because the blue cluster members do not lie along a
sharp sequence but contribute a diminishing fraction relative to the
background at bluer colors.

The green galaxy sample is simply selected to lie between the red and
blue samples defined above, ${(V-i')_{E/SO}}^{+0.1}_{-0.3}$
(Fig.~\ref{colormag}), with generous limits set to include the vast
majority of cluster galaxies, since - as we have established - both
the red and blue samples are negligibly contaminated by cluster
members, and hence the vast majority of cluster members must lie
within this intermediate range of color. A narrow gap on each side of
these samples is left out of our analysis to ensure that the
definition of the background does not encrouch on the cluster
population. Note that unlike the green sample containing the cluster
population, the background populations do not need to be complete in
any sense but should simply be well defined and contain only
background. Increasing the green sample to cover these narrow gaps
does not lead to any particularly significant change in our
conclusions, but only increases somewhat the level of noise by
including relatively more background galaxies. Within the green sample
there are of course background galaxies, and the purpose of this paper
is to make use of the relative proportion of these cluster and
background populations via weak lensing to establish the properties of
the cluster galaxy population, by using the the dilution of the weak
lensing signal of the background galaxies due to the cluster members.

\clearpage
\section{Distortion Analysis of Subaru Images}\label{dist_Sub}

We use the IMCAT package developed by N. Kaiser
\footnote{http://www.ifa.hawaii/kaiser/IMCAT} to perform object
detection, photometry and shape measurements, following the formalism
outlined in \cite*[hereafter KSB]{1995ApJ...449..460K}.
We have modified the method somewhat following the procedures
described in \citet[][see Section 5]{2001A&A...366..717E}.

To obtain an estimate of the reduced shear,
$g_{\alpha}=\gamma_{\alpha}/(1-\kappa)$, 
we measure the image ellipticity $e_{\alpha}$
from the weighted quadrupole moments of the surface brightness of
individual galaxies. 
Firstly the PSF anisotropy needs to be corrected using the star images
as references:
\begin{equation}
e'_{\alpha} = e_{\alpha} - P_{sm}^{\alpha \beta} q^*_{\beta}
\label{eq:qstar}
\end{equation}
where $P_{sm}$ is the {\it smear polarizability} tensor 
being close to diagonal, and
$q^*_{\alpha} = (P_{*sm})^{-1}_{\alpha \beta}e_*^{\beta}$ 
is the stellar anisotropy kernel.
We select bright, unsaturated foreground stars identified in a branch
of the half-light radius ($r_h$) vs. magnitude ($i'$) diagram
($20<i'<22.5$, $\left<r_h\right>_{\rm median}=2.38$ pixels) to
calculate $q^*_{\alpha}$.
In order to obtain a smooth map of $q^*_{\alpha}$ which is used in
equation (\ref{eq:qstar}), we divided the $9{\rm K}\times 7.4{\rm K}$
image into $5\times 4$ chunks each with $1.8{\rm K}\times 1.85{\rm K}$
pixels, and then fitted the $q^*$ in each chunk independently with
second-order bi-polynomials, $q_*^{\alpha}(\vec{\theta})$, in
conjunction with iterative $\sigma$-clipping rejection on each
component of the residual
$e^*_{\alpha}-P_{*sm}^{\alpha\beta}q^*_{\beta}(\vec{\theta})$.  The
final stellar sample consists of 540 stars, or the mean surface number
density of $n_*=0.72$ arcmin$^{-2}$. From the rest of the object
catalog, we select objects with $2.4 \lesssim r_h \lesssim 15$ pixels
as an $i'$-selected weak lensing galaxy sample, which contains $61,115$
galaxies or $\bar{n}_g\simeq 81$ arcmin$^{-2}$.
It is worth
noting that the mean stellar ellipticity before correction is
$(\bar{e_1}^*, \bar{e_2}^*) \simeq (-0.013, -0.018)$ 
over the data field, while the residual
$e^*_{\alpha}$ after correction
is reduced to $ {\bar{e}^{*{\rm res}}_1} = (0.47\pm
1.32)\times 10^{-4}$, $ {\bar{e}^{*{\rm res}}_2} = (0.54\pm
0.94)\times 10^{-4}$.
The mean offset from the null expectation is 
$|\bar{e}^{* \rm res}| = (0.71\pm 1.12) \times 10^{-4}$.
On the other hand, the rms value of stellar ellipticities,
$\sigma_{e*}\equiv\left<|e^*|^2\right>$, is reduced from $2.64\%$ to
$0.38\%$ when applying the anisotropic PSF correction.

Second, we need to correct the isotropic smearing effect on 
image ellipticities
caused by seeing and the window function used for the shape
measurements. The pre-seeing reduced shear $g_\alpha$ can be
estimated from 
\begin{equation}
\label{eq:raw_g}
g_{\alpha} =(P_g^{-1})_{\alpha\beta} e'_{\beta}
\end{equation}
with the {\it pre-seeing shear polarizability} tensor
$P^g_{\alpha\beta}$.
We follow the procedure described in \cite{2001A&A...366..717E} to measure
$P^g$  \cite[see also \S~3.4 of][]{2006astro.ph..6571H}.
We adopt the scalar correction scheme, namely,
\begin{equation}
P^g_{\alpha\beta}=\frac{1}{2}{\rm tr}[P^g]\delta_{\alpha\beta}\equiv
P^g_{\rm s}\delta_{\alpha\beta}
\end{equation}
\citep{1998ApJ...503..531H,1998ApJ...504..636H,2001A&A...366..717E,2006astro.ph..6571H}.
The $P_{g}^{\rm s}$ measured for individual objects are still noisy
especially for small and faint objects.  We thus adopt a smoothing
scheme in object parameter space proposed by \citet[][see also
\citealp{2001A&A...366..717E,2003ApJ...597...98H}]{2000A&A...358...30V}.
We first identify thirty neighbors for each object in
$r_g$-$i'$ parameter space.
We then calculate over the local ensemble
the median value $\langle P_g^{\rm s}\rangle$
of $P_g^{\rm s}$ 
and the variance $\sigma^2_{g}$ 
of $g=g_1+ig_2$ using equation (\ref{eq:raw_g}).
The dispersion $\sigma_g$ is used as an rms error of the shear estimate
for individual galaxies.
The mean variance $\bar{\sigma}_g^2$ over the sample is obtained as
$\simeq 0.171$, or $\sqrt{\bar{\sigma}_g^2}\approx 0.41$.

In the previous study by \cite{2005ApJ...619L.143B}, those objects
that yield a negative value of the raw $P_g^{\rm s}$ estimate were
removed from the final galaxy catalog to avoid noisy shear estimates.
On the other hand, in the present study, we use all of the galaxies in
our weak lensing sample including galaxies with $P_g^{\rm s} <0$.
After smoothing $P_g^{\rm s}$ in the object parameter space, all of
the objects yield positive values of $\left< P_g^{\rm s}\right>$, with
the minimum of $\approx 0.04$.
The median value of $\left< P_g^{\rm s}\right>$ over the weak lensing
galaxy sample, including galaxies with $P_g^{\rm s} <0$, is calculated
as $\approx 0.32$. For a reference, the sub-sample of galaxies with
$P_g^{\rm s}>0$ gives the median average of $\approx 0.33$, mostly
weighted by galaxies with $r_g = 2 - 2.5$ pixels. Finally, we use the
following estimator for the reduced shear:
\begin{equation}
g_{\alpha} = e'_{\alpha}/\left< P_g^{\rm s}\right>.
\end{equation}

The quadratic shape distortion of an object is described by
the complex reduced-shear,  $g=g_1+ig_2$. The tangential
component $g_{T}$ is used to obtain the azimuthally averaged
distortion due to lensing, and computed from the distortion
coefficients $g_{1},g_{2}$:
\begin{equation}
g_{T}=-( g_{1}\cos2\theta +  g_{2}\sin2\theta),
\end{equation}
where $\theta$ is the position angle of an object with respect to the
cluster center, and the uncertainty in the $g_T$ measurement is
$\sigma \equiv \sigma_g/\sqrt{2}$ in terms of the rms error $\sigma_g$
for the complex shear measurement. 
The cluster center is well determined from symmetry of the strong
lensing pattern \citep{2005ApJ...621...53B}. The estimation of $g_{T}$
only has significance when evaluated statistically over large number
of galaxies, since galaxies themselves are not round objects but have
a wide spread in intrinsic shapes and orientations. In radial bins we
calculate the weighted average of the $g_T$s and the weighted error:
\begin{eqnarray}\label{eq:mean_gt}
\langle{}g_{T}(r_n)\rangle{}
&=&\frac{\sum{}g_{T}/\sigma^2}{\sum{}1/\sigma^2}\\
\sigma_T(r_n)
&=& \left(\sum 1/\sigma^2 \right)^{-1/2}.
\end{eqnarray}
\begin{figure}[tb]
  \centering
  \includegraphics[scale=.50]{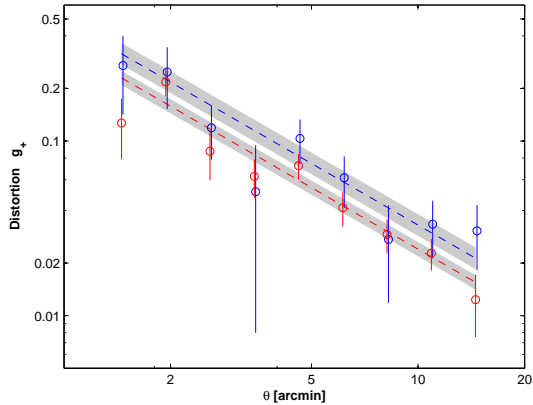}
  \caption{Tangential shear profiles, $g_T(r)$ of the red and blue
    background populations. The tangential distortion profile of both
    decline smoothly from the center, remaining positive to the limit
    of the data. The red galaxies are fitted well with a simple
    power-law, $d\log g_T/d\log r=-1.17\pm0.1$. The blue sample is
    more noisy but also well represented by the same relation, only
    offset in amplitude by $23\pm17\%$ and is related to the greater
    depth of the blue population relative to the red,
    \S~\ref{Photo-z}}
  \label{gt_ofst}
\end{figure}
\begin{figure}[bph]
  \centering
  \includegraphics[scale=.50]{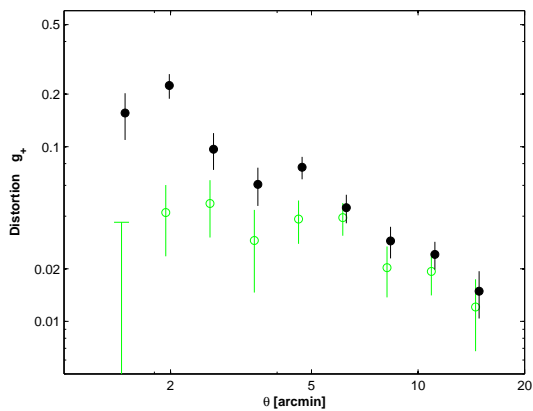}
  \caption{Tangential shear vs. radius. The green points represent the
  intermediate color sample, containing both cluster and background
  galaxies. The black points show the level of tangential distortion
  of the combined red+blue sample of the background.  The green points
  fall close to the background level defined at large radii,
  indicating the green sample is dominated by background galaxies, and 
  falls short towards the cluster center where cluster members increasingly 
  dilute the lensing signal.}
  \label{gt_sub}
\end{figure}

It has been shown that such weights depend on the size of the objects
but mostly on their magnitudes (see e.g. \cite{2006ApJ...647..116H}).
Therefore, as apparent magnitude increases with redshift the redshift
distribution of sources will be modified to some extent by this
weighting scheme. We have investigated this using the catalogue of
\cite{2004AJ....127..180C} as here photometric redshifts are estimated
(see \S~\ref{Photo-z} for a fuller desrciption of the photometric properties
of this sample). First, we generated from the Capak catalog blue/red
background galaxy samples with the same color-magnitude criteria
as the present study. We then derived an i'-magnitude vs. photo-z
relation for each galaxy sub-sample. We subdivided the data into
magnitude (i') bins and derived a magnitude (i') vs. photo-z relation
using median averaging.

We then assume this magnitude-redshift relation holds in our A1689
data and obtain for each galaxy in A1689 an estimate of redshift via
the magnitude -- photo-z relation. It is then straightforward to have
an effective redshift distribution taking into account weak lensing
statistical weights, w. We can see a qualitative feature that although
low-z background galaxies are more strongly weighted than higher-z
ones, the effect on our observed redshift distribution is negligible
because our redshift selection window does not sample these larger
angle, lower redshift objects, but rather the more distant faint
population whose small anglular sizes are heavily influenced by the
seeing.

In Figure~\ref{gt_ofst} we compare the radial profile of $g_T$ of the
red and blue samples defined above. These have a very similar form
implying that the blue sample, like the red sample, is dominated by
background galaxies with negligible dilution by cluster members even
at small radius where the cluster overdensity is large. Very
interestingly a clear offset is visible between these profiles over
the full range of radius, with the amplitude of the blue sample lying
systematically above that of the red sample, as shown in
Figure~\ref{gt_ofst}. This is readily explained as a depth related
effect, as we show below in \S~\ref{Photo-z} where we evaluate the
redshift distributions of these two populations. We find the blue
sample to be deeper than the red, and since lensing scales with depth,
so should the lensing profiles be offset from each other by the same
scale.

The tangential distortion of the green population behaves quite
differently (Fig.~\ref{gt_sub}) falling well below the background
level near the cluster center. The green sample has a maximum signal
at intermediate radius, $3\arcmin-5\arcmin$, and then declines quickly
inside this radius as the unlensed cluster galaxies dominate over the
background in the center. Notice that the green sample does not fall
to zero at the outskirts but rises up to almost meet the level of the
background sample, indicating that the majority of the green sample at
large radius comprises background rather than cluster members. We go
on to use the ratio of the distortion of the green sample compared
with the background level to determine the proportion of cluster
members in \S~\ref{dilution}, but to do so we first evaluate the
expected depths of our samples in \S~\ref{Photo-z}, in order to make a
precise comparison of the lensing signals between them, since the
lensing signal scales geometrically with increasing source distance
and must be accounted for in any comparisons.
 
The results of this paper depend on the ratio of the background
distortion to the cluster contaminated distortion,
$(g_T^{(B)}/g_T^{(G)})$, so that the 5-10\% level calibration
correction factors estimated from simulations done by the STEP project
\citep{2006MNRAS.368.1323H} for the various weak distortion methods
are not of major concern for the bulk of our work.

\section{Distortion Analysis of ACS/HST images}\label{dist_ACS}

In the center of the cluster inside a radius of approximately
$1\arcmin$ the Subaru data become limited in depth by the extended bright 
haloes of the many luminous central galaxies. This region is far better 
resolved and more deeply imaged with HST/ACS in 20 orbits of imaging 
shared between the $g'r'i'z'$ passbands. Many multiple images are known 
here, defining accurately the shape of both the tangential and radial 
critical curves \citep{2005ApJ...621...53B}. Here we analyze the 
statistical distortion of the shapes of the many galaxies recorded 
in these images, to extend our analysis of the properties of the 
cluster galaxies into the center, allowing an accurately defined
central cluster luminosity function to faint luminosities. In
addition, it will be interesting to see how consistent the distortion
profile derived here independently matches the mass profile obtained 
previously from the strongly lensed multiple images.

\begin{figure}[tb]
  \centering \includegraphics[scale=.50]{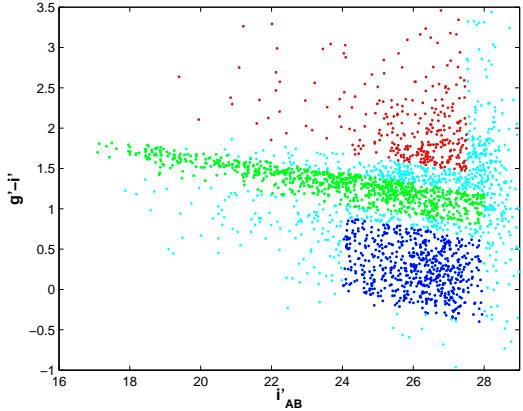}
    \caption{Color vs. magnitude diagram for the central region
    covered by ACS photometry with magnitudes transformed to AB system
    to match the V,I photometry from Subaru. Notice the prominent E/SO
    sequence and the greater depth of 
these data 
compared with Subaru
    shown in Figure 1. A one-to-one comparison of magnitudes for
    objects 
in both datasets 
is shown in Figure~\ref{i_comp}. } \label{colormag_acs}
\end{figure}

We stick to very similar definitions of the three-color 
selected population as with Subaru, but extend their depths by an 
additional $1^m.5$ magnitudes since the ACS data are so much deeper. 
The ACS images are limited to m=28.5 ($5\sigma$)
in each of the passbands. The reduction of the ACS image and the
photometry for the faint sources is described in detail in
\cite{2005ApJ...621...53B}, including the subtraction of the bright
central galaxies in the cluster which is essential for obtaining
accurate photometry and shape measurements of central lensing images
including radial arcs and demagnified central images. For the
distortion analysis we prefer the {\sc im2shape} method developed by
\cite{2002sgdh.conf...38B} for dealing in particular with relatively
elongated images produced by lensing in the strongly lensed region.
This is an improvement over the standard KSB method which we used in
the weak lensing regime appropriate for everything except the central
region $r<2\arcmin$, and used for the Subaru analysis described above.

Using this method, galaxies are fit to a sum of two sheared Gaussians
convolved with a PSF. Each Gaussian has two free parameters, amplitude
and width. The centroid and the shear are also allowed to vary, but
these are restricted to be the same for
both Gaussians. 
Meanwhile, the PSF for each galaxy is determined based on models 
described in \cite{2005ApJ...618...46J}. These PSF models for ACS's WFC 
were derived from observations of the globular cluster 47 Tuc (PROP 9656,
P.I. De Marchi). As the distortion measurements are performed in the
detection image, each galaxy is assigned an "average" PSF based on the
different filters and chip positions in which it was observed.

\begin{figure}[tb]
  \centering
  \includegraphics[scale=.50]{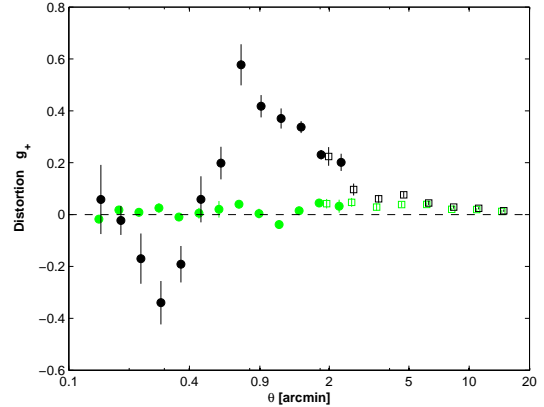}
  \caption{Distortion profile for the central $r<2\arcmin$ area
    covered by ACS (filled circles) together with the weaker
    distortions measured by Subaru at larger radius (open symbols).
    The black points represent the combined blue and red sample of
    background galaxies, and the green points include the cluster
    members. A remarkable saw-tooth pattern is visible for the
    background galaxy distortions in the strong lensing region where
    the tangential and radial critical radii are clearly visible
    corresponding to a maximum and a minimum in the value of $g_T$,
    respectively. In between the distortion passes through zero, where
    the degree of tangential and radial distortion is equal, leaving
    images unchanged in shape at a radius where $\kappa=1$. The
    distortion of the green sample is consistent with zero in the
    inner region where the cluster members dominate the sample but at
    larger radius the green and black points merge for $r>3\arcmin$,
    indicating that there the green sample comprises predominately
    background galaxies.} \label{gt_full}
\end{figure}

We plot the resulting values of $g_T(r)$ (Fig.~\ref{gt_full}) for the
blue, green and red galaxies defined in the same color ranges as the
Subaru data, but to fainter magnitudes. A very well defined
saw-tooth pattern is visible, showing that images are maximally radially
aligned at about $17\arcsec$ and then maximally tangentially aligned at
about $47\arcsec$. This is a very clear signature of strong
lensing, where the maximum corresponds to the location of the 
tangential critical curve (Einstein radius), and the 
minimum to the radial critical curve, where images are maximally
stretched in the radial direction generating a ring of long images 
pointing to the center of mass, as found in \cite{2005ApJ...621...53B}. 
The location of these critical radii agrees very well with 
those derived from the model to the
strong lensing data for this cluster, fitted by
\cite{2005ApJ...621...53B}. 

Another clearly defined radius can also be identified from the point
where the images distortion goes through zero, $g_T=0$, at a radius in
between these two critical radii at about $r\simeq27\arcsec$. It is
important to note that in this region, between the two critical
curves, the parity is $p=-1$ (odd parity), and here 
\be
g=\frac{1}{e^*}=\frac{e}{|e|^2}
\ee \citep[see][]{1995ApJ...439L...1K}.
Here, instead of measuring $g$ we are measuring $-1<e<1$. (This
is different than in the weak lensing region, outside the tangential
critical curve, where $p=+1$ and $g=e$, and therefore no distinction
needed to be made). Since 
\be
e=\frac{1}{g^*}\propto1-\kappa,
\ee
zero distortion corresponds to $\kappa=1$ curve, which lies in between
the tangential and critical curves.

Note, for several reasons we cannot expect that the data
will reach the theoretically extreme value of $g_T=1$ at the
tangential critical radius, meaning that the images are infinitely
stretched tangentially, and also $g_T=-1$ at the radial critical
radius where they are stretched infinitely in the radial direction. By
definition, weak lensing measurements will underestimate the sting
distortions near critical curves. In addition, convolution by the
redshift distribution of the background sources will smooth these
features out. Nonetheless, we can define their locations in radius
rather precisely and these positions must be reproduced in any
satisfactory model, at $r\simeq17\arcsec$ and $r\sim48\arcsec$. In
addition the radius at which $g_T=0$ is also well defined at about
$r\simeq 28\arcsec$ and corresponds to a surface density where
$\kappa=1$, supplying another important constraint on model mass
profiles.

Outside the tangential critical curve we find that the tangential
shear of the background sample (black circles in Fig.~\ref{gt_full})
drops to $g_{T,B}\sim0.2$ at $r=2\arcmin$, in good agreement with the
Subaru analysis at this radius, giving us confidence in the
consistency of our work. For the color range defined above, which
includes the cluster sequence and all the bluer members of the cluster
galaxy population, $g_{T,G}\sim 0$ over the full range of radius of
the ACS data (green circles in Fig.~\ref{gt_full}), indicating - as 
expected - that the galaxy population in the this color range is 
dominated by cluster members with negligible background contamination.

\section{Photometric Redshifts}\label{Photo-z}

We need to estimate the respective depths of our color-magnitude
selected samples when estimating the cluster mass profile, because the
lensing signal increases with source distance, and therefore must
differ between the samples. The effect of this difference in distance
on the weak lensing signal is simply linear as we can see from the
relation between the dimensionless surface mass density,
\be
\kappa(r)=\Sigma(r)/\Sigma_{crit}, 
\ee
where
\be
\Sigma_{crit}=\frac{c^2}{4\pi Gd_l}\frac{d_s}{d_{ls}}
\ee
and the tangential distortion:
\be
\langle g_T(r)\rangle=(\bar{\kappa}(r)-\kappa(r))/(1-\kappa(r))
\ee 
so that in the weak limit where $\kappa$ is small, 
\be\label{eq:gt_D}
\langle g_T(r)\rangle \propto  {{d_{ls}\over{d_s}}{(\bar{\Sigma}(r)-\Sigma(r))}}
\ee
and hence for an individual cluster, with a fixed redshift and a given
mass profile, the observed level of the weak distortion simply scales
with the lensing distance ratio. Further details are presented in the 
appendix.
%
The mean ratio of $d_{ls}/d_s$, which is weighted by the redshift 
distribution of the background population corresponding to our magnitude 
and color cuts, is calculated using the expression 
\be\label{eq:meanD}
\langle D\rangle\equiv\langle{d_{ls}\over{d_s}}\rangle=\frac{\int{\displaystyle{d_{ls}\over{d_s}}(z)N(z)dz}}{\int{N(z)dz}} .
\ee

Since we cannot derive complete samples of reliable photometric
redshifts from our limited 2-color V,i' images of A1689, we instead
make use of other deep field photometry covering a wider range of
passbands, sufficient for photometric redshift estimation of the faint
field redshift distribution appropriate for samples with the same
color and magnitude limits as our red, green and blue populations.

The photometry of \cite{2004AJ....127..180C} is very well suited for
our purposes, consisting of relatively deep multi-color photometry
over a wide field taken with Subaru, producing reliable photometric
redshifts for the majority of field galaxies to faint limiting
magnitudes. The \cite{2004AJ....127..180C} galaxy catalog contains 
almost 50,000 galaxies over $0.2$ sq. deg. with $UBVRIZ$ photometry. 
We have estimated photometric redshifts for this catalog using the 
Bayesian based method of \cite{2000ApJ...536..571B}, with a prior 
based on the redshift and spectral type distributions of the HDF-N, 
with a spectral library containing the templates of 
\cite{2004ApJS..150....1B} with an additional two blue starburst
galaxies as described in \cite{2006AJ....132..926C}.
  
\begin{figure}[t]
  \centering
  \includegraphics[scale=.50]{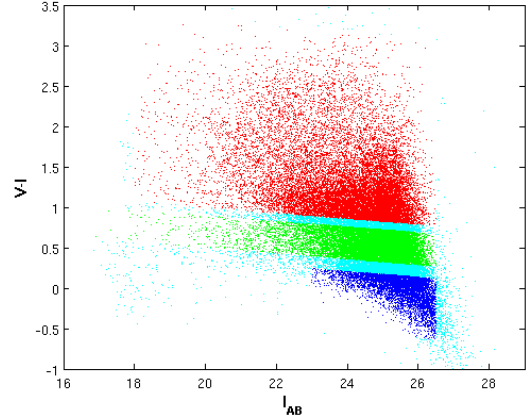}
  \caption{ Color--magnitude diagram for Capak galaxy 
    catalog for the same passbands as the photometry of A1689 shown in
    Fig~\ref{colormag}. This photometry is of a field region and
    serves as our reference for evaluating the expected depth of the
    background samples defined in \S~\ref{data}.}
  \label{colormag_capak}
\end{figure}

A full redshift probability distribution is produced for each galaxy
of the form: 
\be\label{bpz} 
p(z|C)\propto \sum_T p(z,T|m_0)p(C|z,T), 
\ee 
where $p(C|z,T)$ 
is the redshift likelihood obtained by comparing the observed colors 
$C$ with the redshifted library of templates $T$. The factor $p(z,T|m_0)$ 
is a prior which represents the redshift/spectral mix distribution as a 
function of the observed $I-$band magnitude. We use a prior which 
describes the redshift/spectral type mix in the HDF-N, which has been 
shown to significantly reduce the number of ``catastrophic'' errors 
($\Delta z >1$) in the photometric redshift catalog \cite[see][and 
references therein]{2004ApJS..150....1B}. For each galaxy we look at 
its redshift probability distribution p(z) and identify up to 3 redshift 
local maxima. Each of these maxima corresponds to a redshift $z_i$, 
spectral type $t_i$, and a discretized probability $p_i(z_i,t_i) \leq 1$. 
Using this information we generate a mock observation of all the $z_i,t_i$
combinations in the Subaru filters, and then build a redshift
histogram by selecting galaxies using the same color cuts and adding
up their probabilities in each redshift bin.
\begin{figure}[b]
  \centering \includegraphics[scale=.50]{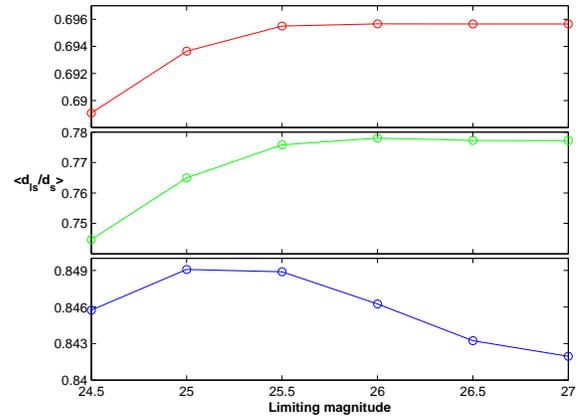}
  \caption{The weighted mean lensing depth $d_{ls}/d_s$ as a function
    of the apparent i'-magnitude limit of the red, green \& blue
    backgrounds, calculated using the photometric redshifts of the
    \cite{2004AJ....127..180C} sample.  The expected depth of the
    samples differs significantly between the samples and in general the
    distance ratio grows only slowly with increasing apparent
    magnitude for each sample.}
  \label{Dmean}
\end{figure}
\\
\begin{figure}[b]
  \centering \includegraphics[scale=.50]{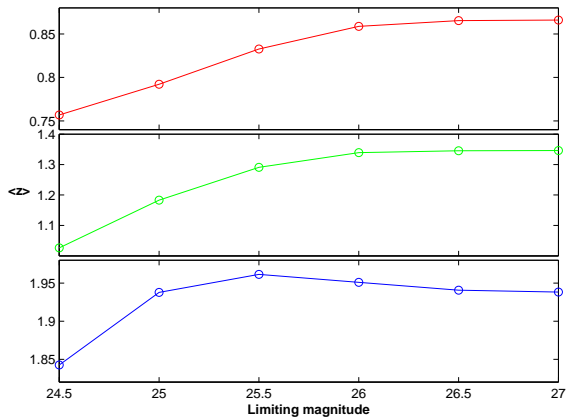}
    \caption{Similar to the previous figure but the redshift is
      plotted against apparent magnitude 
      corresponding to the mean depth calculated above. The average
      redshift differs significantly between the three samples, being
      lowest for the red background $\langle z\rangle\sim0.85$ and
      highest for the blue $\langle z\rangle\sim2$.}
    \label{Zmean}
\bigskip
\end{figure}

The color-magnitude diagram for the Capak catalog galaxies
is shown in Figure~\ref{colormag_capak}, where the equivalent
color-magnitude selected samples are displayed. The resulting mean
redshift of the background galaxies in each of our three
color-selected samples is caclulated as a function of limiting
magnuite of the sample (Fig.~\ref{Zmean}), by using the redshift
distribution from the Capak catalog. The redshift distribution is also
used to evaluate the weighted mean depths $\langle D\rangle$ (shown in
Fig.~\ref{Dmean} as a function of sample limiting magnitude), for
comparing the weak lensing amplitudes between the green and the
red+blue samples.  
This is done by dividing up each sample into 81
independent bins of $2\arcmin$, calculating the weighted mean redshift
and depth in each, and taking the mean value and variance over the
bins. The mean redshift of the red sample is only $\langle
z_{red}\rangle\sim0.871\pm0.045$, whereas the blue sample is
calculated to have, $\langle z_{blue}\rangle\sim2.012\pm0.124$.  The
green sample lies in between with, $\langle
z_{green}\rangle\sim1.429\pm0.093$. The weighted relative depths of
these samples using equation (\ref{eq:meanD}), for samples selected to
our magnitude limit of $i<26.5$, are $\langle
D_{red}\rangle=0.693\pm0.012$, $\langle
D_{green}\rangle=0.728\pm0.012$, and $\langle D_{blue}\rangle=
0.830\pm0.011$, and the corresponding redshifts $z_D$ equivalent to
these mean depths are, $z_{D,red}=0.68$, $z_{D,green}=0.79$,
$z_{D,blue}=1.53$, respectively. Hence the ratio of the mean depth of
the blue sample to the red sample is $\langle D_{blue}\rangle/\langle
D_{red}\rangle=1.20$, accounting well for the observed offset seen in
Figure~\ref{gt_ofst}.

We also make use of the Capak ``green'' sample to investigate the
level of ``cosmic variance'' in $d_{ls}/d_s$, and although there is
variation in the redshift distribution the variance of the mean
redshift is remarkably tight, and as quoted above we find a very small
variance associated with the mean lensing depth,
$\sigma(\langle{}d_{ls}/d_s\rangle{})=0.015$. This stability is also a
feature noticed in pencil beam redshift surveys in general, that the
mean depth is stable to spikes in the redshift distribution, e.g.,
\cite{1988MNRAS.235..827B}.

The form of the distance ratio $D$ can be expressed in terms of 
the redshifts of the source and lens for a given set of cosmological 
parameters. In the main case of interest, that of a flat model with a 
nonzero cosmological constant, the relation is given by 
\begin{eqnarray} 
\label{Dz}
{d_{ls}\over{d_s}}& =& 1-\frac{(1+z_l)^2 \zeta(z_l)}{(1+z_s)^2 
\zeta(z_s)}\\
\zeta (x) & = & \int^x_0 \frac{dz}{[\Omega_\Lambda+
\Omega_M(1+z)^3]^{1/2}} .
\end{eqnarray}
General expressions for the dependence of this distance ratio on
arbitrary combinations of $\Omega_M$ and $\Omega_\Lambda$ are lengthy
and can be found in \cite{1990MNRAS.246P..24F}. For a low redshift
cluster like A1689 (z=0.183), the form of this function is rather flat
for sources at $z>1$, see \cite{2005ApJ...621...53B}. Therefore the
main uncertainty in determining the cosmological parameters from a
comparison of $g_T$ between samples of different redshifts, is small
compared with clustering noise along a given line of sight behind the
cluster, as examined in detail by \citep{2005ApJ...621...53B}.  Thus,
we do not seriously examine this effect here but rather simply adopt
the recent (three year) WMAP cosmological parameters
\citep{2006astro.ph..3449S} when making the above depth correction.
With sufficient number of clusters and similar or better photometric
redshift information (from multiple filter observations) one can hope
to examine the trend of redshift vs. lensing distance in the future.

Note that lensing magnification, $\mu$, will modify $g_T$ slightly by
increasing the depth to a fixed magnitude. But the magnification is
small, $\mu<0.^{m}2$ over most of the cluster, $r>3'$. In any case, the
dependence of the mean redshift on depth is a slow function of
redshift, so that it is safe for our main purposes to ignore the
effect of magnification on the depth of our samples. Furthermore,
since we are only interested in the proportion of the cluster relative
to the background for our purposes, we are not affected by the
modification of the background number counts caused by lensing, which
has been shown to significantly deplete the surface density of
background red galaxies in A1689, and found to be consistent with the
predicted level of magnification based on the distortion measurements
\citep{2005ApJ...619L.143B}.

\section{Weak Lensing Dilution}\label{dilution}

We can now estimate the number density of cluster galaxies by taking the
ratio of the weak lensing signal between the green sample and the
background sample, with the background including both red and blue 
galaxies selected from the Subaru catalog as 
explained in \S~\ref{data}, and accounting for differences in the 
relative depths of these samples, as explained in \S~\ref{Photo-z}.

\begin{figure}[tb]
  \centering \includegraphics[scale=.50]{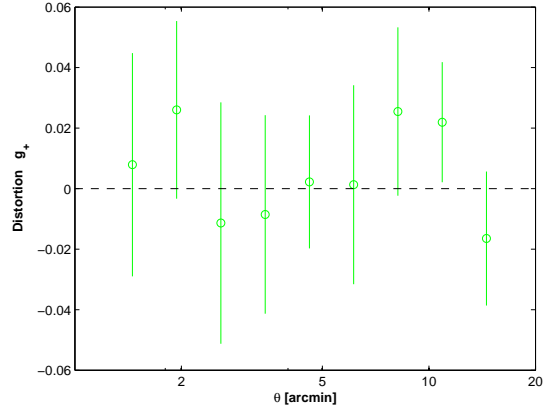}
  \caption{Tangential distortion of the bright cluster
    sequence ($i'<21^m.5$) galaxies plotted against radius from the
    cluster center. By choosing the bright part of the sequence we
    minimize the background contamination, and can therefore check that
    the tangential distortion of the cluster members is negligible,
    which indeed is very clear from this figure.}
  \label{gt_green}
\end{figure}

Cluster members are unlensed and hence assumed to have random
orientations, so they are expected to contribute no net tangential
lensing signal. This assumption can be examined for the brighter
($i'<21^m.5$) cluster sequence galaxies whose tight sequence protrudes
beyond the faint field background (Fig.~\ref{colormag}) with
negligible background contamination, so that we are secure in
selecting this subset to test the assumption that the cluster galaxies
are randomly oriented.  Indeed, the net tangential signal of this
population is consistent with zero (Fig.~\ref{gt_green}),
$g_T^{(G)}=0.0043\pm0.0091$.

\begin{figure}[tb]
  \centering \includegraphics[scale=.50]{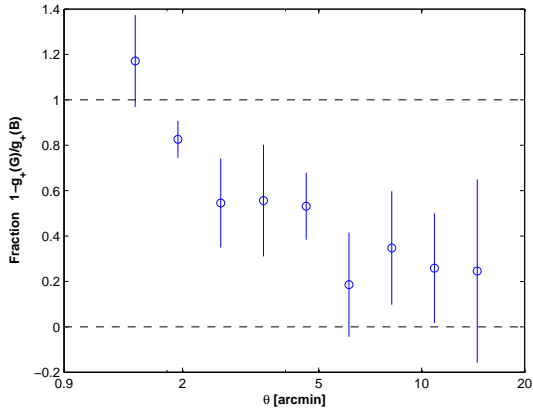}
  \caption{Fraction of cluster membership vs. radius. 
    Cluster membership is proportional to the dilution of the distortion
    signal of the green sample, relative to the expected distortion of
    the background galaxies set by the red and blue samples. As
    expected, close to the cluster center the fraction of cluster
    members in the green sample becomes maximal, whereas at larger
    radius $r>3\arcmin$, the fraction of cluster members is small,
    indicating that most of the green sample comprises background
    galaxies.}
  \label{frac}
\smallskip
\end{figure}

For a given radial bin ($r_n$) containing objects in the green sample,
whose width in color has been chosen to encompass the full range of
cluster galaxies (\S~\ref{data}), the mean value of $g_T^{(G)}$
(eq.~\ref{eq:mean_gt}) is an average over background and cluster
members. Thus, its mean value 
$\langle g_T^{(G)}\rangle$ will be lower than the true background level 
denoted by $\langle g_T^{(B)}\rangle$ (Fig.~\ref{gt_sub}) in proportion 
to the fraction of unlensed galaxies in the bin that lie in the cluster 
(rather than in the background), 
since the cluster members on average will add no net tangential
signal. Therefore,
\begin{equation}\label{eq:frac}
f_{cl}(r_n)\equiv\frac{N_{cl}}{N_{Green}}
=1-\frac{\langle{}g_T(r_n)^{(G)}\rangle{}}{\langle{}g_T(r_n)^{(B)}\rangle{}}\frac{\langle
  D^{(B)}\rangle}{\langle D^{(G)}\rangle}
\end{equation}
is the cluster membership fraction of the green sample (see full
derivation in the appendix).
 
Thus, we can use this effect to quantify statistically the 
number of cluster galaxies 
by comparing $g_T^{(G)}$ with the true background level derived from the 
pure background red and blue samples $g_T^{(B)}$, at a fixed radius. 
This is shown in Figure~\ref{frac}, where we have also taken into account 
the effect of the relative depths of the differing samples. We find that 
the fraction of cluster members drops smoothly from $\sim100\%$ within 
$r<2\arcmin$ to only $\sim20\%$ at the limit of the data, 
$r\sim 15\arcmin$.

\section{Cluster light and color profiles}\label{lum_sec}

To determine the luminosity profile of the cluster galaxies, we need
to go further, because in general the brightness distribution of the
cluster members is different than that of the background galaxies; 
specifically, it is skewed to brighter magnitudes, certainly for the 
bulk of the cluster sequence. To account generally for any difference 
in the brightness distributions we can subtract a ``$g_T$-weighted'' 
luminosity contribution of each galaxy, which when averaged over the 
distribution will have zero contribution from the unlensed cluster 
members. We first calculate the ``g-weighted'' correction in arbitrary 
flux units. We estimate the total flux of the cluster in the $n^{th}$
radial bin,
\begin{equation}
F_{cl}(r_n)=\displaystyle\sum_{i}{F_i^{(G)}}
-\frac{\langle{}D^{(B)}\rangle{}/\langle{}D^{(G)}\rangle{}}{\langle{}g_T(r_n)^{(B)}\rangle{}}
\displaystyle\sum_{i}{F_i^{(G)}\cdot{}g_{T,i}^{(G)}}
\end{equation}
where the sum is over all galaxies in the radial bin.

The flux is then translated back to apparent magnitude, and from that
the luminosity is derived. First we calculate the absolute magnitude,
\begin{equation}
M_{i'}=i'-5\log{d_L}-K(z)+5,
\end{equation}
where the $K$-correction is evaluated for each radial bin according to
its $V-i'$ color, 
which - after the correction is made for each of the bands - is now 
the cluster color. The luminosity is then
\begin{equation}
L_{i'}=10^{0.4(M_{i'\odot}-M_{i'})}L_{i'\odot},
\end{equation}
where $M_{i'\odot}=4.54$ is the absolute $i'$ magnitude of the Sun (AB
system).

\begin{figure}[tb]
  \centering \includegraphics[scale=.50]{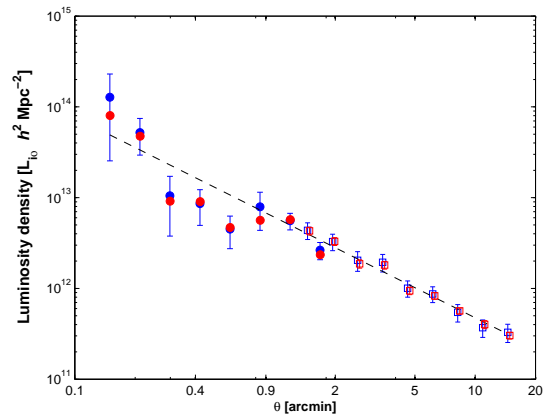}
  \caption{The ``g-Weighted'' luminosity density vs. cluster
    radius. Each galaxy's flux $F_i$ (green sample) is weighted by its
    tangential distortion $g_{T,i}$ with respect to the background
    distortion signal (red points). The filled circles represent the
    ACS data, and the empty squares are for the Subaru data. The blue
    points are derived from integrating over the luminosity functions
    of the same radial bins (see \S~\ref{lum_func}), and serve as a
    consistency check, showing good agreement between these differing
    calculations. The dashed line is the best fitting linear relation
    for the blue points.}
  \label{lum_profile}
\end{figure}

\begin{figure}[bt]
  \centering
  \includegraphics[scale=.50]{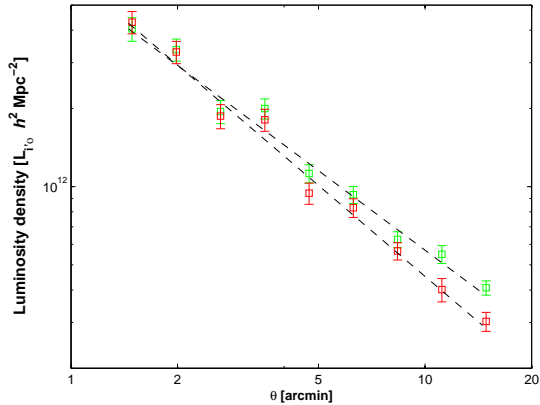}
  \caption{The ``g-Weighted'' luminosity density vs. cluster
    radius (red squares), compared to the unweighted luminosity
    density (green circles), showing as expected the increasing size
    of the correction with increasing radius, where the sample becomes
    increasingly dominated by background galaxies.}
  \label{lum_p_comp}
\end{figure}

The result yields the luminosity profile of the cluster as shown in
Figure~\ref{lum_profile} (red squares). Here we can see that the
cluster luminosity profile is well approximated by a simple power-law
with a projected slope of $d\log (L_{i'})/d\log(r)\sim -1.12\pm
0.06$, to the limit of the data. We also show in
Figure~\ref{lum_p_comp} the unweighted luminosity profile with no
correction for the field, demonstrating that the g-weighted correction
is negligible at small radius as expected, since the cluster dominates
numerically over the background, but becomes increasingly more
important at larger radius where the background dominates. Note that
we derive a more accurate inner luminosity profile using the ACS
photometry for the central region (Fig.~\ref{lum_profile}, red
circles), and here there is only a negligible correction for the
background due to the high central density of galaxies in this
cluster.

In a careful study of clusters and groups identified in the SDSS
survey, \cite{2005ApJ...633..122H} find a similar slope for the most
massive clusters, in terms of the composite surface density profile of
$d\log(n)/d\log(r)\simeq -1.05\pm0.04$, over the radius range
$r<2$Mpc, with slightly shallower slopes occurring in the less
overdense clusters and groups. This may be compared directly with our
slope derived above, assuming a constant $M_*/L$, for the ratio of
galaxy mass to galaxy luminosity. More directly we derive a density
profile using the
$f_{cl} n^{(G)}$ 
which also gives a slope of $\simeq-0.9\pm0.09$.


\begin{figure}[tb]
  \centering \includegraphics[scale=.50]{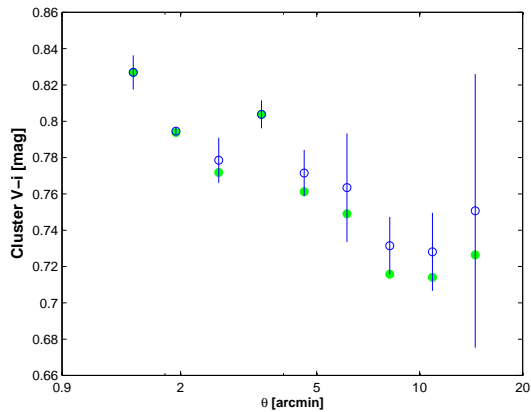}
  \caption{Galaxy color profile after weighting the
    color of each object by its individual distortion, $g_i$,
    accounting for any difference between the color distribution of
    the cluster and background populations comprising the green galaxy
    population. The color of the cluster members becomes slowly bluer
    with increasing radius moving from E/SO colors in the center to
    mid-type galaxy colors at the limit of the data, $r\sim
    2~h^{-1}$Mpc. The green points represent the uncorrected $V-i'$
    profile of the green sample. }
  \label{color_C}
\end{figure}

In the same manner, we construct a ``g-weighted'' color profile of the
cluster, $V-i'$, which we have corrected as described above. We obtain
a color profile that shows a weak tendency towards bluer colors with
increasing radius, as expected, indicating a tendency towards
later-type galaxies at large radius (Fig.~\ref{color_C}). Also shown
is the unweighted color profile (green points), which again is
steeper due to the uncorrected field component which dominates
numerically over the cluster at larger radius and is generally bluer
in color than the cluster. This change in color with radius
corresponds to a significant radial gradient in spectral type, from
predominantly early-type with $(V-i')_{AB}\simeq 0.84$, to mid sequence
type, Sb, with $(V-i')_{AB}\simeq 0.73$, and indicates that for this
cluster very blue starburst and Scd galaxies are not the dominant
population at the limiting radius of our sample ($r\sim 2~h^{-1}$Mpc),
where otherwise the color would tend to $(V-i')_{AB}\simeq 0.5$, using
standard template sets \citep{2004ApJS..150....1B}. We go on to make
use of this color-radius relation in \S~\ref{ML_sec}, when examining
the radial profile of the ratio of total cluster mass to the stellar
mass in galaxies. We do this by correcting the luminosity profile for
the tendency towards more luminous early-type stars that are
responsible for the bluer galaxy colors at large radius and which
otherwise bias the interpretation of the $M/L$ ratio, as described below.

\section{Cluster Luminosity Functions}\label{lum_func}

\begin{figure}[tb]
  \centering \includegraphics[scale=.50]{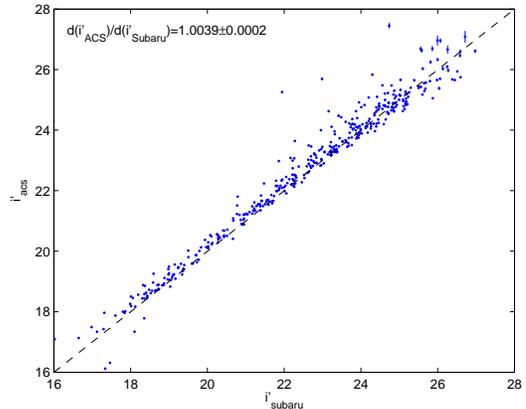}
  \caption{Comparison between ACS and Subaru photometry for objects in 
  common in the central region covered by both datasets, $r<2\arcmin$. 
  There is very good agreement between magnitudes of the two datasets 
  which have independent zero-points and of course independeent photometry.}
  \label{i_comp}
\end{figure}

The data allow the luminosity function to be usefully constructed in
several independent radial and magnitude bins, and hence we can
examine the form of the luminosity function of cluster members as a
function of projected distance from the cluster center. For this we
combine the ACS and the Subaru photometry. The ACS has the advantage
of extending two magnitudes fainter in the $i'$-band than the Subaru
photometry for $r<2\arcmin$. As can be seen in Figure~\ref{i_comp},
the ACS $i'$ magnitudes agree with Subaru $i'$ magnitudes for galaxies
found and matched in both 
catalogs.

\begin{figure*}[thb]
  \centering \includegraphics[scale=0.65]{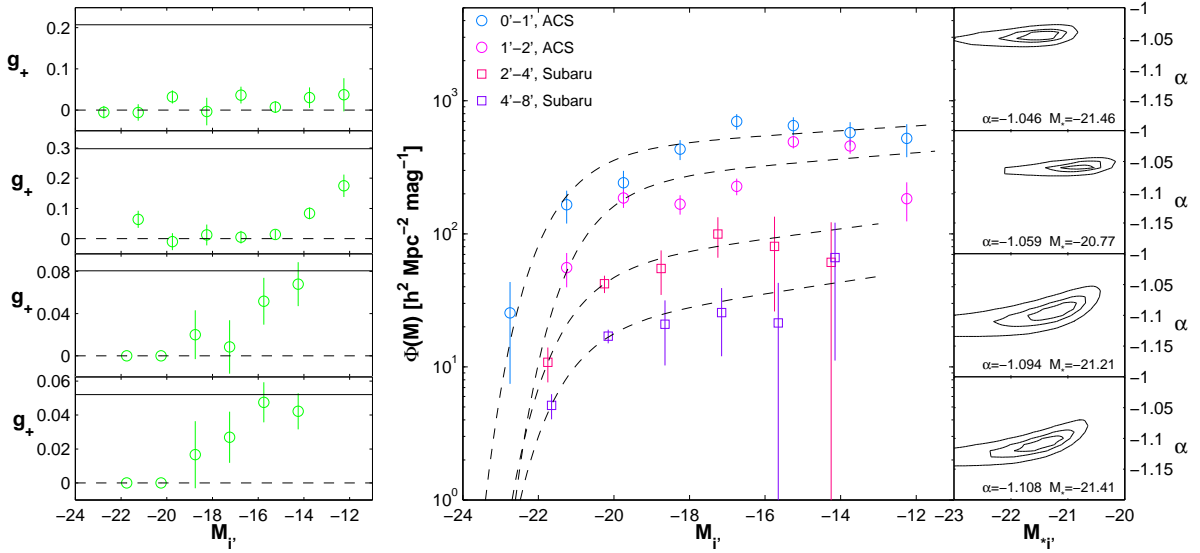}
  \caption{Center: Luminosity functions are shown for four independent
  radial bins, indicating little trend with radius. The left-hand panel
  is the degree of tangential distortion $g_T$ as a function of
  magnitude used in the derivation of the corresponding luminosity
  function. Notice that in general $g_T$ increases with decreasing
  luminosity because the level of background increases at fainter
  magnitudes. The right-hand panel shows the $1\sigma,~2\sigma$, and
  $3\sigma$ contours for the Schechter function parameters $M_*$ and
  $\alpha$, for each of the corresponding luminosity functions.}
  \label{LF}
\end{figure*}

The background correction to evaluate 
$f_{cl}$, 
must be made in each
magnitude bin independently, since the relative proportion of
background galaxies increases with apparent magnitude, so that the
lower luminosity bins of the green luminosity function are expected to
contain a greater fraction of background galaxies and hence should have
a relatively higher value of $g_T^{(G)}$. This trend is apparent in
Figure~\ref{LF} (left panels), where we plot the recovered mean
tangential distortion (here the average is over a magnitude bin) for
each of the four radial bins, as a function of absolute magnitude. A
clear trend is found at all radii towards higher levels of $g_T$ at
fainter luminosities. Note that the mean level of the background
distortion (black solid line) drops with increasing radius so that the
proportion $g_T^{(G)}(M)/g_T^{(B)}$ is generally an increasing function of
radius and a decreasing function of luminosity. To correct for this we
simply apply equation~(\ref{eq:frac}) to each 
magnitude bin:
\begin{equation}
\Phi_{cl}(M_k)=\Phi(M_k)\cdot[1-\langle{}g_T^{(G)}(M_k)\rangle{}/\langle{}g_T^{(B)}(r)\rangle{}]
\end{equation}
(Note that the background signal is averaged over the whole range of 
magnitudes at that radius.) 

\begin{figure}[thb]
  \centering \includegraphics[scale=0.5]{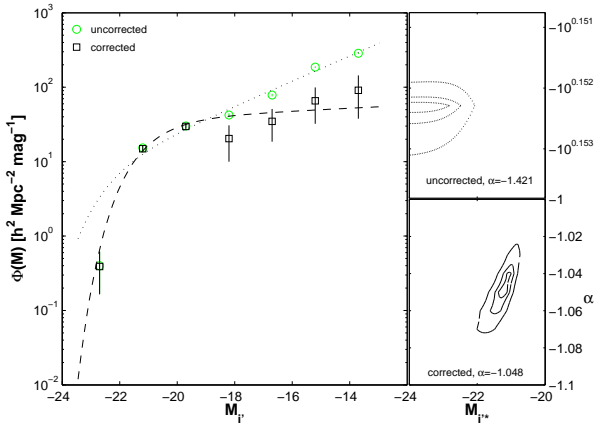}
  \caption{Composite luminosity function of the cluster (black
  squares), for $r<10\arcmin$, with the fit to a Schechter function
  displayed in the lower right panel. The green circles show the
  luminosity distribution without the correction made for the
  background dilution. At bright magnitudes, there is little
  difference between the uncorrected and the corrected points,
  however, at the faint end the uncorrected distribution rises, which
  if uncorrected would overestimate the faint end slope
  $\alpha\sim1.4$ (top right panel), showing clearly the magnitude of
  the background correction, which when accounted for by our method
  results a flat faint end slope, $\alpha\sim1$.}  \label{LF_comp}
\end{figure}

We then construct the luminosity function for four independent radial
bins, as shown in Figure~\ref{LF} (middle panel) and fit a
\cite{1976ApJ...203..297S} function to each (dashed lines). It can be
seen that there is no obvious tendency for the shape of the luminosity
function to change with radius. The faint-end slope of a Schechter
function fit is $\alpha= -1.05\pm0.07$ in the $i'$-band. This constancy
with radius has been argued with somewhat less significance in other
well studied massive clusters \citep[e.g.,][]{2005MNRAS.364.1147P},
based on similar deep 2-color imaging, where the limiting radius is
more restricted.
We also construct a composite luminosity function for the whole
cluster (Fig.\ref{LF_comp}), for $r<10\arcmin$, which shows clearly
the effect of our ``g-weighted'' background correction, without which
the faint-end slope would be considerably steeper, $\alpha\sim1.4$.

Our approach is of course essentially free of uncertainties in the
subtraction of background galaxies by its nature.  While qualitative
similarity between the results of the various studies is clear,
agreement in detail is not necessarily expected, given the likely
dispersion in the strength of this effect between clusters. Also, the
question of background contamination is always an issue in the
standard approach due to the inherent fluctuations in the surface
density of background galaxies, and the need to establish the
background counts at a sufficiently large radius from the cluster to
avoid self-subtraction of the cluster at the boundaries of the data, a
subject explored in depth, e.g. \cite{2000A&A...353..930A,2001A&A...367...59P,2005MNRAS.360..727A,2005ApJ...633..122H,2005A&A...433..415P}.

We also integrate our luminosity functions as a consistency check of
the luminosity density profiles derived earlier. This is done by 
calculating $\Phi_{cl}(M)$ in the same radial bins as our luminosity 
profile above, and summing over the magnitude bins:
\begin{equation}
L_{cl}(r_n)
=\displaystyle\sum_{k}\Phi_{cl}(M_k)\cdot\Delta
M_k\cdot 10^{0.4(M_{i'\odot}-M_k)}
\end{equation}
The results shown above in Figure~\ref{lum_profile} (blue points) agree 
very well with those of $L_{cl}$ described in the previous section.

Note that in 
constructing these luminosity density profiles we have
implicitly assumed that the luminosity function is integrated over
fully. Fortunately our data is complete to a sufficient depth
($i'<26.5$) so that the contribution of the integrated luminosity
density from undetected objects is very small, as evaluated when we
examine the luminosity functions. The difference between integrating
up to a limiting magnitude of $M_{i'}<-14$ and extrapolating up to
$M_{i'}<-10$ is only about $0.1\%$.

The lack of any obvious upturn in the cluster luminosity function to
very faint luminosities, $M_{i'}<-12$, in the cluster core, is in
agreement with several other studies based on deep photometry of large
cluster samples, e.g., the composite cluster luminosity function derived
by \cite{1997AJ....113..117G,1999A&A...342..408G,2001A&A...367...59P,
2005ApJ...633..122H,2005MNRAS.364.1147P}, where wide field imaging is 
employed for several Abell clusters and
overdensities identified in the SDSS survey, and where careful
attention can be paid to the background counts and their uncertainty.
The study of \cite{2005MNRAS.364.1147P} is the most similar to our own,
containing three rich and fairly distant Abell clusters, and here the
LF's show no obvious upturn to $M<-13$ with a generally flat Schechter
function slope in the range -1.1 to -1.25.

For the well studied Coma cluster, a steeper slope has been claimed,
$\alpha\simeq-1.4$, by \cite{1995AJ....110.1507B}, though subsequent
faint spectroscopy by \cite{2000A&A...353..930A}, has revealed the
presence of a background cluster at $z\simeq 0.5$, which when
corrected for leads to a flat faint-end slope. In contrast an upturn
is claimed for a composite sample of 25 SDSS selected clusters by
\cite{2005A&A...433..415P}, though an individual examination shows
considerable variation, with only a minority of $\sim 6$ displaying a
distinct upturn which varies in amplitude, so that one may wonder
about the role of anomalous background count fluctuations in these
cases.

\section{$M/L$ profiles}\label{ML_sec}

\begin{figure}[tb]
  \centering
  \includegraphics[scale=.50]{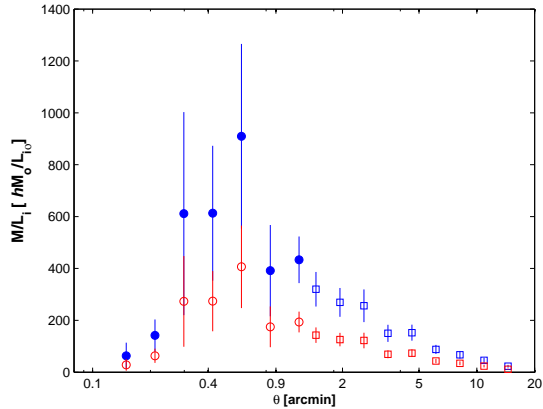}
  \caption{Mass-to-light ratio vs. radius. Mass profile is taken
    from a fit made by \cite{2005ApJ...619L.143B} using the same data
    analyzed here (blue points). We also plot the total
    mass-to-stellar mass ratio $M/M_{*}$ (red points), accounting for
    the cluster color profile.  }
  \label{ML}
\end{figure}

We may now go on to examine the mass-to-light ratio using the mass
profile previously determined based on the distortion and
magnification profile of the red galaxy sample, as derived by
\cite{2005ApJ...619L.143B}, and also based on the central strong
lensing information derived from 106 multiple-images
\citep{2005ApJ...621...53B}.  The mass profile derived in
\cite{2005ApJ...619L.143B} was found to be somewhat more pronounced
than a simple NFW form, with the observed gradient increasing
monotonically with radius from the cluster center.  Dividing the mass
profile by our newly derived light profile we obtain a profile of the
mass-to-light ratio for A1689 as shown in Figure~\ref{ML}.

\begin{figure}[tb]
  \centering
  \includegraphics[scale=.50]{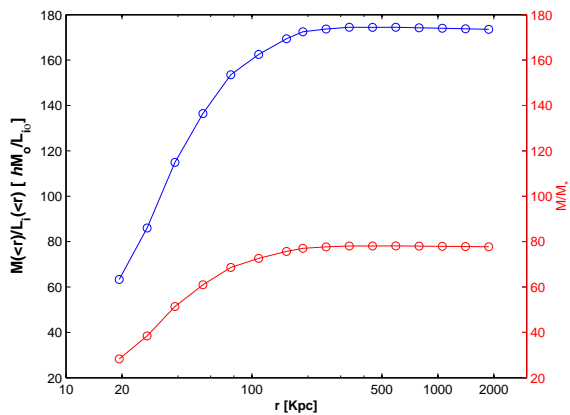}
  \caption{Cumulative Mass-to-light ratio vs. radius. The profile
    declines markedly inside $r<100~h^{-1}$kpc, otherwise it remains
    fairly constant with radius.}
  \label{MLcum}
\smallskip
\end{figure}

We find that the $M/L$ ratio peaks at intermediate radius around
$2\arcmin$ ($r\sim100~h^{-1}$Kpc) and then falls off linearly to
larger radius.  \citep{2005ApJ...621...53B} have previously identified
the drop in $M/L$ at small radius as due principally to the tight
central clump of very luminous cluster members, noting that 
this clump may have resulted from the effect of dynamical friction.

Note that the peak value is rather large, equivalent to
$M/L_B\sim 400h(M/L)_{\odot}$ in the restframe B-band often used as a
reference, but the mass of A1689 is at the extreme end of the cluster
population, $M\sim2 \times 10^{15}hM_{\odot}$
\citep{2005ApJ...619L.143B}, and given the general tendency of $M/L$
to increase with increasing mass, from galaxies through groups to
clusters, we may not be surprised to find the peak to somewhat exceed
the typical range for clusters, $150-300h{M/L_B}_{\odot}$
\citep{2001ApJ...552..427C}. The general profile of $M/L$ is similar 
in form to that derived for CL 0152-1357 by \cite{2005ApJ...618...46J}
based on a careful weak lensing analysis of recent deep ACS images.

We have also constructed a profile of the total mass to stellar mass
$M/M_*$ ratio. This is arguably a more physically useful indicator of
the relationship between dark and luminous matter compared to the
ratio of $M/L$, because the starlight can be strongly influenced by
the presence of relatively small numbers of luminous hot stars. To
calculate $M/M_*$ we make use of the color profile derived in
\S~\ref{lum_sec}, and an empirical relationship between color and the
ratio $M_*/L$ for stellar populations established for local galaxies
in the the SDSS survey by \cite{2003ApJS..149..289B}. The slope of the
projected stellar mass profile, $d\log M_*/d\log r = -1.15\pm0.13$,
derived this way is slightly steeper than the luminosity profile, as
expected. The observed relation we derive this way is somewhat flatter
than for $M/L$ and the mean contribution by mass for stars is about
$1.25\%$ for this cluster and similar to a mean value of $\sim 2\%$
derived from a carefully selected sample of local clusters by
\cite{2006A&A...452...75B}.

\section{Cluster Mass Profile}\label{MLconsist}

\begin{figure}[t]
  \centering \includegraphics[scale=.50]{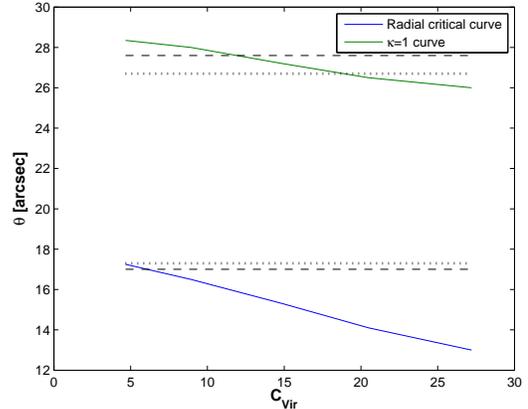} 
  \caption{The curves shows how the radius of the radial critical
    curve, shown in blue, varies with the concentration parameter of
    an NFW profile, where the model is normalized to generate the
    observed Einstein radius of $47\arcsec$ for all values of the
    concentration parameter, $C_{vir}$. The radial critical curve
    shrinks as the mass profile becomes steeper. This is also the case
    for the radius at which the distortion $g_T=0$, corresponding to
    the radius where the surface density of matter is equal to the
    critical surface density ($\kappa=1$), here the degree of
    tangential and radial distortion is always equal, independent of
    the form of the mass profile. We have marked the observed values
    of the radial critical curves and the radius where the distortion
    is seen to be zero as measured independently in two ways using the
    statistical distortion measurements (dotted lines) as described in
    \S~\ref{dist_ACS}, and the same values derived from the multiple
    image analysis presented in \cite{2005ApJ...621...53B} (dashed
    lines).  These differing measurements are consistent with each
    other and by comparison with the model curves bracket intermediate
    values of concentration in the range $5<C_{vir}<15$ for the inner
    strongly lensed region. }
\label{curves}
\end{figure}

We use the combined distortion information obtained from the ACS and
Subaru imaging, as described above (Fig.~\ref{gt_full}) and compare
with models for the mass distribution. We have improved on our earlier
distortion measurements made with Subaru, with the addition of the
background blue galaxy population defined here, so that the
significance of the distortion measurements is somewhat greater than
our earlier work which was based only on the red sample
\citep{2005ApJ...619L.143B}. In addition, we have extended the
distortion measurements to the central region using the HST/ACS
information, as described in \S~\ref{dist_ACS}, where we have clearly
identified a maximum and a minimum value of $g_T$, which accurately
correspond to the tangential and radial critical curves
(Fig.\ref{curves}), independently derived from the the many giant
tangential and radial arcs observed for this cluster
\cite[see][]{2005ApJ...621...53B}.

Here we test the universal parameterization of CDM-based mass profiles
advocated by \cite{1997ApJ...490..493N}. This model profile is
weighted over the differing results from sets of haloes identified in
N-body simulations. A cluster profile is summed over all the mass
contained within the main halo, including the galactic haloes. Hence,
we compare the integrated mass profile we deduced directly with the
NFW predictions without having to invent a prescription to remove the
cluster galaxies.

NFW have shown that massive CDM haloes are predicted to be less
concentrated with increasing halo mass, a trend identified with
collapse redshift, which is generally higher for smaller haloes
following from the steep evolution of the cosmological density of
matter. The most massive bound structures form later in hierarchical
models and therefore clusters are anticipated to have a relatively low
concentration, quantified by the ratio $C_{vir}=r_{virial}/r_s$. In
the context of this model, the predicted form of CDM dominated are
predicted to follow a density profile lacking a core, but with a much
shallower central profile ($r\leq 100~h^{-1}$ kpc) than a purely
isothermal body.

The fit to an NFW profile is made keeping $r_s$, and $\rho_s$, the 
characteristic radius and the corresponding density, as free parameters. 
These can be adjusted to normalize the model to the observed maximum
in the distortion profile at the tangential critical radius of 
$\simeq 45\arcsec$. The combination of these parameters then fixes the 
degree of concentration, and the corresponding lensing distortion 
profile can then be calculated.

Integrating
the mass along a column, z, where $r^2=({\xi_r}{r_s})^2+$z$^2$ gives:
\be\label{Mcol}
{M}(\xi)={\rho_s}{r_s^3}(\xi)\int_o^{\xi}d^2\xi
\int_{-\infty}^{\infty} {\frac{1}{(r/r_s)(1+r/r_s)^2}}{\rm{dz}\over{r_s}}.
\ee
Using this mass, a bend-angle of 
$\alpha=\frac{4GM(<\theta)}{c^2\theta{d_l}} \frac{d_{ls}}{d_s}$ is
produced at position 
${\theta}=\xi r_s /d_l$.
The mean interior mass within some radius 
$r_x=x c_{x} r_s$ can be obtained by integration of the NFW profile
giving: 
\be\label{McolNFW}
M(<r)={4\pi\over{3}}r_x^3{\Omega_m(o)\Delta_c(1+z)^3\over{x^3\Omega_m(zl)}}
{\left[{\log(1+xc_x)-{xc_x\over{1+xc_x}}\over{\log(1+c_x)-{c_x\over{1+c_x}}}}\right]}, 
\ee 
where the above integral is carried out to the virial radius. Here
when we make this comparison we do not attempt to fit the distortions
in the region of the radial and tangential critical curves because the
measurements must underestimate the amplitude of the model predictions
near these curves due to the finite size of the background galaxies,
so that the model maximum and minimum, $g_T=\pm1$ corresponding to
the tangential and radial critical curves cannot be reached by the data
(see Fig.~\ref{gt_full}) near these critical curves. Finite area
sources are on average not as magnified or distorted as an ideal point
source due to the gradient of the lensing magnification over the
surface of the source, so that for a source lying over a lens caustic
only an infinitesimally small area is infinitely magnified. In
principle simulations based on realistic galaxy samples like those
modeled by \cite{1998ApJ...506..557B} could be used to correct for
this effect but this will await further work.

We can instead utilize here the observed location of these critical
curves since a clear maximum and minimum is observed in the inner
distortion data (Fig.~\ref{gt_full}) and has also been independently
determined from the multiple-image data \citep{2005ApJ...621...53B}
for this cluster. In addition to the location of the critical curves
we can also clearly identify the radius where $g_T=0$, lying in
between these critical curves. At this radius the radial and
tangential magnifications are equal and hence images are unchanged in
shape (though in general highly magnified), so the observed value of
$g_T$ will pass through zero at a radius in between the critical
curves. This radius corresponds to the contour where the projected
surface density is equal to the critical surface density,
$\Sigma=\Sigma_{crit}$ \citep[e.g.,][]{1995ApJ...439L...1K}, and hence
is smaller for more concentrated profiles.  In Figure~\ref{curves} we
plot these two radii as a function of model concentration, where the
models are all normalized to a tangential critical radius of
$47\arcsec$ to match the mean critical radius derived from the data
\citep{2005ApJ...621...53B}.

To normalize the models we choose to reproduce the observed Einstein
radius of $47\arcsec$ and compare the predicted location of the radial
critical curve and the $\kappa=1$ curve. Figure~\ref{curves} shows
that both these radii decrease slowly as the concentration parameter
is increased. We have marked the observed values of these radii as
determined by two independent observational means. We can use the
statistical distortion measurements as described in \S~\ref{dist_Sub},
and the same values derived from the multiple image analysis presented
in \cite{2005ApJ...621...53B}. These differing estimates are closely
consistent with each other, and by comparison with the model curves
bracket intermediate values of concentration in the range
$5<C_{vir}<15$ (Fig.~\ref{curves}), a range consistent with the
results from a detailed fit to the inner profile measured in
\cite{2005ApJ...621...53B}. This is found to be very similar to the
independently derived central profile of A1689 from
\cite{2005MNRAS.362.1247D,2006ApJ...640..639Z,2006astro.ph..5470H}.

\begin{figure}[t]
  \centering \includegraphics[scale=.50]{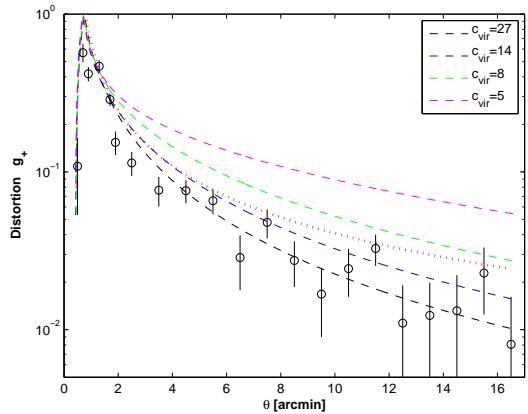} 
  \caption{NFW models are compared with the measured values of $g_T$ for the
    red+blue background sample described in \S~\ref{data}. The models
    are normalized to match the observed Einstein radius of
    $47\arcsec$. A relatively high concentration is preferred with the
    best fit corresponding to $C_{vir}=27^{+3.5}_{-5.7}$ (black
    curve), for the bulk of data, $r>1.5\arcmin$, excluding the strong
    region where the measured distortions required a significant
    correction for the finite source sizes. The anticipated low
    concentration of $C\sim 5$ (pink curve) is obviously excluded by
    the data. For reference we also overplot the purely isothermal
    profile normalized to the observed Einstein radius (red dotted
    curve). It also overpredicts the data. }
  \label{gfit}
\end{figure}

Outside the tangential critical radius, for $r>1.5$, the distortion
measurements are small enough not to suffer from any significant
underestimation due to the finite source sizes and we may compare the
observed distortion profile, $g_T(r)$, out to the limit of our data,
$r<15\arcsec$. We find this distortion profile is reasonably well
fitted by an NFW profile particularly at large radius,
$4.0\arcsec<r<15\arcsec$, but with a relatively large concentration,
$C_{vir}=27.2^{+3.5}_{-5.7}$, as shown in Figure~\ref{gfit}. Note that
we have used here a linear radial binning when measuring the
concentration parameter and therefore the result here is more weighted
by large radius signal than for the analysis of
\cite{2005ApJ...619L.143B} where we used logarithmic binning, yielding
a smaller value of $C_{vir}$. This difference in the derived value of
$C_{vir}$ is not becuase of any revision in our estimates of the
distortion, in fact both analyses yield very consistent distortion
profiles at large radius, but rather that the form of the NFW profile
is not consistent with our data over the full radial range - the best
fitting NFW model is either too shallow at large radius or too steep
at small radius depending where one prefers to fit the data.

We also plot lower concentration profiles, including $C_{vir}=14$,
which was found previously by \cite{2005ApJ...621...53B}, to fit best
the overall lensing derived mass profile from combining the mass
profile derived from the multiply lensed images in the central region,
$r<2\arcmin$, with the mass distribution derived from weak lensing
distortion and magnification measurements from the red background
galaxy sample. This model fit, as pointed out by
\cite{2005ApJ...619L.143B}, is not as pronounced as the observed
surface mass profile, being too shallow at larger radius and too steep
at small radius \cite[see figures 1\&3 of][]{2005ApJ...619L.143B}.
Here we see more clearly that this fit with $C_{vir}=14$ increasingly
overpredicts the observed distortion profile with radius. We also plot
$C_{vir}=8.2$ which best fits the central strongly lensed region
\citep{2005ApJ...621...53B} $r<2\arcmin$, derived from 106 multiply
lensed images. Again this fit overpredicts the $g_T$ profile in the
weak lensing regime, as pointed out in \cite{2005ApJ...619L.143B}.

We clearly exclude the low concentration profile generally predicted
by CDM based models of structure formation. A value of $C_{vir}\sim 5$
is generally anticipated for massive clusters, although the scatter in
concentration at a given mass is considerable
\cite[e.g.,][]{2001MNRAS.321..559B}. Figure~\ref{gfit} shows clearly
how this profile is much too shallow to generate the relatively
steeply declining observed distortion profile. The triaxiality of
realistic haloes means that projection effects will bias somewhat the
derived distortion profile, as examined carefully by
\cite{2005ApJ...632..841O} and \cite{2005astro.ph..6171H}, showing
that the level of such bias effect is expected to enhance the derived
concentration by approximately $\sim 20\%$ on average. Whilst A1689 is
clearly an anomalous cluster in terms of the size of the Einstein
radius, the cluster is very round in terms of the projected X-ray
emission, with only minimal substructure observed in the optical near
the center. Hence we are left with a clearly unresolved problem, that
the observed concentration would seem to far exceed any reasonable
estimate. Other independent work on the combined profile from strong
and weak lensing measurements for the clusters Cl0024+17
\citep{2003ApJ...598..804K} and MS2137-23 \citep{2003A&A...403...11G}
also point to surprisingly high concentrations, and it is therefore
important to extend this type of detailed work to other clusters to
test the generality of the profile derived here.

For reference we also plot the distortion profile for a singular
isothermal body in Figure~\ref{gfit}, which is simply expressed as
\begin{equation}
g_T=\frac{1}{2\frac{\theta}{\theta_E}-1},
\end{equation}
and normalized to the observed Einstein radius, $\theta_E=47\arcsec$.
This model also overpredicts the data at large radius, indicating the
outer mass profile is steeper than $1/\theta$ in projection.

\section{Discussion and Conclusions}\label{discussion}

We have explored a new approach to deriving the luminous properties of
cluster galaxies by utilizing lensing distortion measurements, based
on the dilution of the lensing distortion signal by unlensed cluster
members which we assume are randomly oriented. We have tested this
assumption for a restricted sample of bright cluster galaxies which
project beyond the faint galaxy background population so the level of
background contamination is negligible, confirming that the cluster
galaxies are randomly oriented with a negligible net tangential
distortion for the purposes of our work.

This dilution approach is applied to A1689 to derive the radial light
profile of the cluster, a color profile and radial luminosity
functions. The light profile is found to be smoothly declining and
fitted with a power-law slope $d\log(L)/d\log(r)=-1.12\pm0.06$. We
also see a mild color gradient corresponding to a change in the
cluster population from early- to mid-type galaxies in moving from the
center out to the limit of our data at 2~$h^{-1}$Mpc. Unlike the light
profile the gradient of mass profile is continuously steepening, such
that the ratio of $M/L$ peaks at intermediate radius.  
We find that the
cluster luminosity function has a flat faint-end slope of
$\alpha=-1.05\pm0.07$, nearly independent of radius and with no faint
upturn to $M_{i'}<{-12}$.

A major advantage of our approach is that we do not need to define
far-field counts for subtracting a background, as in the usual method
where there is a limitation imposed by the clustering of the
background population that limits the radius to which a reliable
subtraction of the background can be made.

We have also established that the bluest galaxies in the field of A1689 lie
predominantly in the background, as their radial distortion profile
follows closely the red galaxies, but with an offset indicating the
blue population lies at a greater mean distance than the red
background galaxies, and consistent with the estimated mean redshifts
of these two populations. With a larger sample of clusters, this
purely geometric effect can potentially be put to use to provide a
simple model-independent measure of the cosmological curvature.

The mass profile of A1689 was reexamined using our combined background
sample of red and blue galaxies. The distortion profile derived from
this sample is consistent with our earlier work, but somewhat more
statistically significant, so we have examined the mass profile more
carefully out to a larger radius. We have found that the distortion
profile is steeper than predicted for CDM haloes appropriate for
cluster sized masses $C_{vir}\sim 5$. This discrepancy is particularly
clear at large radius $r>2\arcmin$, where an acceptable fit is found
to an NFW profile but with a concentration $C_{vir}=27^{+3.5}_{-5.7}$.
This finding is consistent with our earlier work which showed that
although an overall best fit profile of $C_{vir}\simeq 14$ to the
joint strong and weak lensing based data presented in
\cite{2005ApJ...619L.143B}, the curvature of the data is more
pronounced than an NFW profile being shallower in the inner region and
steeper at larger radius, so that the derived value of the
concentration increases with radius depending on the radial limits
being examined.

This result is surprising and may require a significant departure from
the standard CDM model, either in terms of the mass content, or the
epoch at which the bulk of the cluster was assembled. For example, one
possibility to achieve earlier formation of clusters is to allow
deviation from Gaussianity of the primordial density fluctuation
field, as has been considered recently by, e.g.,
\cite{2006MNRAS.368.1583S}. A1689 is amongst the most massive known
clusters, and projection effects may play a role in boosting somewhat
the lensing signal along the line of sight. We therefore aim to test
the generality of this result with a careful study of a statistical
sample of clusters.

Upcoming spatially resolved SZ measurements will add a significant new
ability to determine cluster mass profiles over a large range of
radius, and allow for improved consistency checks between the various
independent means of estimating masses. The combination of X-ray,
lensing and SZ measurements will soon lead to far greater accuracy in
understanding the nature of cluster mass profiles.

We plan an improvement to the weak lensing work with deeper
multi-color imaging from Subaru for measuring reliable photometric
redshifts for a sizable fraction of the background population. This
added dimension of depth will enhance the weak lensing signal and
reduce the systematic problems of cluster and foreground contamination
of the lensing signal. We also aim to extend this work to well studied
clusters at lower redshift with archived Subaru imaging and detailed
X-ray and upcoming SZ observations as in principle, the lensing signal
should be equally strong for lower redshift clusters, given the
maximal ratio of lens to source distances, $d_{ls}/d_s\simeq 1$, for
faint background sources.

\acknowledgments We are grateful to N.~Kaiser for making the IMCAT
package publicly available. We thank Masahiro Takada, Doron Lemze and
Eran Ofek for useful discussions and in addition we thank Eran Ofek
for useful Matlab routines. We thank our referee Raphael Gavazzi for a
thorough reading of the manuscript and many useful suggestions. TJB
thanks the generosity of the JSPS and the hospitality of NAOJ and the
Astronomical Institute of Tohoku University. ACS was developed under
NASA contract NAS5-32865, and this research was supported in part by
NASA grant NAG5-7697. Work at Tel-Aviv University was supported by
Israel Science Foundation grant 214/02.

\begin{appendix}

\section{Appendix A: Cluster Galaxy Fraction from the Lensing Dilution Effect}
\label{appA}

Let us derive eq.~(\ref{eq:frac}).
For simplicity, here we assume the weak lensing limit so that
the reduced shear is approximated by the
gravitational shear, $g\approx \gamma$.
It is useful to factorize the lensing signal with the
geometry-dependent factor such that \citep{1997A&A...318..687S}:
\be 
g_T(r)=w(z)g_{T,\infty}(r)
\ee
where $g_{T,\infty}(r)$ denotes the tangential shear
calculated for
hypothetical sources at an infinite redshift, and $w(z)$ is the
lensing strength of a source at $z$ relative to a source at
$z\rightarrow\infty$, $w(z)=D(z)/D(z\to \infty)$; $D(z)\equiv
d_{ls}/d_s$ as introduced in \S 5. The relative lensing strength 
vanishes for cluster and
foreground galaxies, that is, $w(z)=0$ for $z\le z_l$.

As the tangential shear is obtained by averaging over an annular
region, it can be formally written in the following form:
\be\label{eq:g_D} \langle g_T(r)\rangle = g_{T,\infty}(r) \frac{\int
  d^2x~dz~dn/dz~w(z)}{\int d^2x~dz~dn/dz} = g_{T,\infty}(r)
\frac{\int_{z_l}^{\infty} dz~dN/dz~w(z)}{N_{tot}} = g_{T,\infty}(r)
\frac{N_{bg}}{N_{tot}} \langle w \rangle_{z>z_l} \ee where $dn/dz$ is
the surface number density distribution of galaxies per unit redshift
interval per steradian, $dN/dz= \int d^2x~dn/dz$ is the mean redshift
distribution of galaxies in the annulus,
$N_{tot}=\int_0^{\infty}\!dz~dN/dz$ is the total number of galaxies in
the annulus, $N_{bg}=\int_{z_l}^{\infty}\!dz~dN/dz$ is the number of
background galaxies in the annulus, $\langle w
\rangle_{z>z_l}=\int_{z_l}^{\infty}\!dz\,dN/dz~w(z)/\int_{z_l}^{\infty}\!dz\,dN/dz$
is the mean lensing strength without including the dilution effect;
here we have assumed that the lensing properties are constant over the
annulus where we take the ensemble averaging. Note that the factor
$N_{bg}/N_{tot}$ accounts for the dilution effect on the lensing
signal strength due to contamination by foreground and cluster-member
galaxies. In general, there is a contribution from foreground galaxies
to the total number of galaxies $N_{tot}$. However, for the case of
A1689 at a low redshift of $z_l=0.183$, this contribution is
negligible. That is, $N_{\rm tot}\approx N_{cl}+N_{bg}$ with $N_{cl}$
being the number of cluster galaxies in the annulus. For a background
galaxy sample, $N_{tot}=N_{bg}$.

Since we are to compare galaxy samples with different redshift distributions,
we need to account for different values of the mean lensing strength,
$\langle w\rangle_{z>z_l}$.
As explained above, 
our green sample (denoted with G) comprises both cluster and
background galaxies.
Hence, 
according to eq.~(\ref{eq:g_D}), the expectation value for
the mean tangential shear estimate is 
\be
\langle g_T^{(G)}(r)\rangle =
g_{T,\infty}(r)
\frac{N_{bg}^{(G)}}{N_{cl}+N_{bg}^{(G)}} \langle w^{(G)} \rangle_{z>z_l}.
\ee
As for our background sample (denoted with B), 
including the red and blue samples, 
this is 
\be
\langle g_T^{(B)}(r)\rangle =
g_{T,\infty}(r) \langle w^{(B)} \rangle_{z>z_l}.
\ee
By taking the ratio of the two tangential shear estimates, 
we obtain the following expression:
\be 
 \langle g_T^{(G)}(r)\rangle / 
 \langle g_T^{(B)}(r)\rangle =
\frac{N_{bg}^{(G)}}
     {N_{cl}+N_{bg}^{(G)}} 
\frac{
  \langle w^{(G)} \rangle_{z>z_l}
}
{
  \langle w^{(B)} \rangle_{z>z_l}
}.
\ee

Alternatively, we have the expression for the cluster galaxy fraction
as
\be
\label{eq:fraction_of_cluster_galaxies}
f_{cl}(r)\equiv
\frac{N_{cl}}{N_{cl}+N_{bg}^{(G)}}
=1-
\frac{\langle g_T^{(G)}(r)\rangle}
     {\langle g_T^{(B)}(r)\rangle}
\,
\frac{\langle w^{(B)}\rangle_{z>z_l}}
     {\langle w^{(G)}\rangle_{z>z_l}}
=1-
\frac{\langle g_T^{(G)}(r)\rangle}
     {\langle g_T^{(B)}(r)\rangle}
\,
\frac{\langle D^{(B)}\rangle_{z>z_l}}
     {\langle D^{(G)}\rangle_{z>z_l}}.
\ee
This is the desired formula for the cluster galaxy fraction from the
weak lensing dilution effect. In order to take into account different
populations of background galaxies in the two samples, one needs to
estimate the correction factor, $\langle D^{(B)}\rangle_{z>z_l} /
\langle D^{(G)} \rangle_{z>z_l}$.

\section{Appendix B: Non-linear effect in the reduced shear estimate}\label{appB}

In Appendix A, we assume that the observable reduced shear is
linearly proportional to the lensing strength factor, $w(z)$.
However, the reduced sear, defined as $g=\gamma/(1-\kappa)$, 
is non-linear in $\kappa$, 
so that  the averaging operator with respect to the redshift
generally acts non-linearly on the redshift-dependent components in $g$.

To see this effect, we expand the reduced shear with respect to the
convergence $\kappa$ as
\begin{equation}
g=\gamma(1-\kappa)^{-1}
=w\gamma_{\infty}\,(1-w \kappa_{\infty})^{-1}
=w\gamma_{\infty}\,\sum_{k=0}^{\infty}
\left( 
w\kappa_{\infty}
\right)^{k}
\end{equation}
where $\kappa_{\infty}$ and $\gamma_{\infty}$ are the lensing
convergence and the gravitational shear, respectively, calculated for
a hypothetical source at an infinite redshift.  Hence, the reduced
shear averaged over the source redshift distribution is expressed as
\begin{equation}
\label{eq:mean_g}
\langle g \rangle = 
\gamma_{\infty}\,\sum_{k=0}^{\infty} \langle w^{k+1}\rangle
\kappa_{\infty}^k.
\end{equation}
In the weak lensing limit where $\kappa_{\infty}, |\gamma|_{\infty}
\ll 1$, then $\langle g\rangle \approx \langle w \rangle
\gamma_{\infty}$.  Thus, the mean reduced shear is simply proportional
to the mean lensing strength, $\langle w\rangle$.
The next higher-order approximation 
for eq.~(\ref{eq:mean_g}) is given by
\begin{equation}
\label{eq:2ndapprox}
\langle g\rangle \approx \gamma_{\infty}
\left( \langle w\rangle + \langle w^2\rangle \kappa_{\infty}^2 \right)
\approx \frac{\langle w\rangle \gamma_{\infty}}{1-\kappa_{\infty}\langle
w^2\rangle/\langle w\rangle }.
\end{equation}
\cite{1997A&A...318..687S} found that eq.~(\ref{eq:2ndapprox}) yields
an excellent approximation in the mildly non-linear regime of
$\kappa_{\infty} \lesssim 0.6$.  Defining $f_w\equiv \langle
w^2\rangle/\langle w\rangle^2$, we have the following expression for
the mean reduced shear valid in the mildly non-linear regime:
\begin{equation}
\langle g\rangle \approx 
\frac{\langle \gamma\rangle}
{1-f_w\langle \kappa\rangle}
\end{equation}
with $\langle \kappa\rangle =\langle w\rangle \kappa_{\infty}$ and
$\langle \gamma\rangle =\langle w\rangle \gamma_{\infty}$
\citep{1997A&A...318..687S}.  For lensing clusters located at low
redshifts of $z_l \lesssim 0.2$, $\langle w^2\rangle \simeq \langle w
\rangle^2$ or $f_w\approx 1$, so that $\langle g\rangle \approx
\langle\gamma \rangle/(1-\langle \kappa \rangle)$.

The ratio of tangential shear estimates using 
two different populations B and G
of background galaxies, in the mildly non-linear regime, is given as
\begin{equation}
\langle g_T^{(G)}\rangle/
\langle g_T^{(B)}\rangle \approx \frac{\langle w^{(G)}\rangle}
{\langle w^{(B)} \rangle} \,
\frac{1-f_w^{(B)}\langle w^{(B)}\rangle \kappa_{\infty}}
     {1-f_w^{(G)}\langle w^{(G)}\rangle \kappa_{\infty}}
\approx
\frac{\langle w^{(G)}\rangle}{\langle w^{(B)} \rangle} 
\,
\left\{
  1-( f_w^{(B)} \langle w^{(B)}\rangle 
     -f_w^{(G)} \langle w^{(G)}\rangle)
 \kappa_{\infty}
+O(\langle\kappa\rangle^2)
\right\}.
\end{equation} 
The lowest-order correction term is proportional to
$\left(
      f_w^{(B)} \langle w^{(B)}\rangle 
     -f_w^{(G)} \langle w^{(G)}\rangle
\right)\kappa_{\infty}$, which is much smaller than unity 
for the galaxy samples of our concern in the mildly non-linear regime. 
In conclusion, it is therefore a fair
approximation to use eq.~(\ref{eq:frac})
for measuring the cluster galaxy fraction via the dilution effect.

\end{appendix}

\bibliography{A1689}

\begin{thebibliography}{66}
\expandafter\ifx\csname natexlab\endcsname\relax\def\natexlab#1{#1}\fi

\bibitem[{{Adami} {et~al.}(2000){Adami}, {Ulmer}, {Durret}, {Nichol}, {Mazure},
  {Holden}, {Romer}, \& {Savine}}]{2000A&A...353..930A}
{Adami}, C., {Ulmer}, M.~P., {Durret}, F., {Nichol}, R.~C., {Mazure}, A.,
  {Holden}, B.~P., {Romer}, A.~K., \& {Savine}, C. 2000, \aap, 353, 930

\bibitem[{{Allen}(1998)}]{1998MNRAS.296..392A}
{Allen}, S.~W. 1998, \mnras, 296, 392

\bibitem[{{Andreon} {et~al.}(2005){Andreon}, {Punzi}, \&
  {Grado}}]{2005MNRAS.360..727A}
{Andreon}, S., {Punzi}, G., \& {Grado}, A. 2005, \mnras, 360, 727

\bibitem[{{Arabadjis} {et~al.}(2002){Arabadjis}, {Bautz}, \&
  {Garmire}}]{2002ApJ...572...66A}
{Arabadjis}, J.~S., {Bautz}, M.~W., \& {Garmire}, G.~P. 2002, \apj, 572, 66

\bibitem[{{Bardeau} {et~al.}(2005){Bardeau}, {Kneib}, {Czoske}, {Soucail},
  {Smail}, {Ebeling}, \& {Smith}}]{2005A&A...434..433B}
{Bardeau}, S., {Kneib}, J.-P., {Czoske}, O., {Soucail}, G., {Smail}, I.,
  {Ebeling}, H., \& {Smith}, G.~P. 2005, \aap, 434, 433

\bibitem[{{Bell} {et~al.}(2003){Bell}, {McIntosh}, {Katz}, \&
  {Weinberg}}]{2003ApJS..149..289B}
{Bell}, E.~F., {McIntosh}, D.~H., {Katz}, N., \& {Weinberg}, M.~D. 2003, \apjs,
  149, 289

\bibitem[{{Ben{\'{\i}}tez}(2000)}]{2000ApJ...536..571B}
{Ben{\'{\i}}tez}, N. 2000, \apj, 536, 571

\bibitem[{{Ben{\'{\i}}tez} {et~al.}(2004){Ben{\'{\i}}tez}, {Ford}, {Bouwens},
  {Menanteau}, {Blakeslee}, {Gronwall}, {Illingworth}, {Meurer}, {Broadhurst},
  {Clampin}, {Franx}, {Hartig}, {Magee}, {Sirianni}, {Ardila}, {Bartko},
  {Brown}, {Burrows}, {Cheng}, {Cross}, {Feldman}, {Golimowski}, {Infante},
  {Kimble}, {Krist}, {Lesser}, {Levay}, {Martel}, {Miley}, {Postman}, {Rosati},
  {Sparks}, {Tran}, {Tsvetanov}, {White}, \& {Zheng}}]{2004ApJS..150....1B}
{Ben{\'{\i}}tez}, N. {et~al.} 2004, \apjs, 150, 1

\bibitem[{{Bernstein} {et~al.}(1995){Bernstein}, {Nichol}, {Tyson}, {Ulmer}, \&
  {Wittman}}]{1995AJ....110.1507B}
{Bernstein}, G.~M., {Nichol}, R.~C., {Tyson}, J.~A., {Ulmer}, M.~P., \&
  {Wittman}, D. 1995, \aj, 110, 1507

\bibitem[{{Bertin} \& {Arnouts}(1996)}]{1996A&AS..117..393B}
{Bertin}, E., \& {Arnouts}, S. 1996, \aaps, 117, 393

\bibitem[{{Biviano} \& {Salucci}(2006)}]{2006A&A...452...75B}
{Biviano}, A., \& {Salucci}, P. 2006, \aap, 452, 75

\bibitem[{{Bouwens} {et~al.}(1998){Bouwens}, {Broadhurst}, \&
  {Silk}}]{1998ApJ...506..557B}
{Bouwens}, R., {Broadhurst}, T., \& {Silk}, J. 1998, \apj, 506, 557

\bibitem[{{Bridle} {et~al.}(2002){Bridle}, {Kneib}, {Bardeau}, \&
  {Gull}}]{2002sgdh.conf...38B}
{Bridle}, S., {Kneib}, J.-P., {Bardeau}, S., \& {Gull}, S. 2002, in The shapes
  of galaxies and their dark halos, Proceedings of the Yale Cosmology Workshop
  ''The Shapes of Galaxies and Their Dark Matter Halos'', New Haven,
  Connecticut, USA, 28-30 May 2001. Edited by Priyamvada Natarajan. Singapore:
  World Scientific, 2002, ISBN 9810248482, p.38, ed. P.~{Natarajan}, 38--+

\bibitem[{{Broadhurst} {et~al.}(2005{\natexlab{a}}){Broadhurst},
  {Ben{\'{\i}}tez}, {Coe}, {Sharon}, {Zekser}, {White}, {Ford}, {Bouwens},
  {Blakeslee}, {Clampin}, {Cross}, {Franx}, {Frye}, {Hartig}, {Illingworth},
  {Infante}, {Menanteau}, {Meurer}, {Postman}, {Ardila}, {Bartko}, {Brown},
  {Burrows}, {Cheng}, {Feldman}, {Golimowski}, {Goto}, {Gronwall}, {Herranz},
  {Holden}, {Homeier}, {Krist}, {Lesser}, {Martel}, {Miley}, {Rosati},
  {Sirianni}, {Sparks}, {Steindling}, {Tran}, {Tsvetanov}, \&
  {Zheng}}]{2005ApJ...621...53B}
{Broadhurst}, T. {et~al.} 2005{\natexlab{a}}, \apj, 621, 53

\bibitem[{{Broadhurst} {et~al.}(2005{\natexlab{b}}){Broadhurst}, {Takada},
  {Umetsu}, {Kong}, {Arimoto}, {Chiba}, \& {Futamase}}]{2005ApJ...619L.143B}
{Broadhurst}, T., {Takada}, M., {Umetsu}, K., {Kong}, X., {Arimoto}, N.,
  {Chiba}, M., \& {Futamase}, T. 2005{\natexlab{b}}, \apjl, 619, L143

\bibitem[{{Broadhurst} {et~al.}(1988){Broadhurst}, {Ellis}, \&
  {Shanks}}]{1988MNRAS.235..827B}
{Broadhurst}, T.~J., {Ellis}, R.~S., \& {Shanks}, T. 1988, \mnras, 235, 827

\bibitem[{{Bullock} {et~al.}(2001){Bullock}, {Kolatt}, {Sigad}, {Somerville},
  {Kravtsov}, {Klypin}, {Primack}, \& {Dekel}}]{2001MNRAS.321..559B}
{Bullock}, J.~S., {Kolatt}, T.~S., {Sigad}, Y., {Somerville}, R.~S.,
  {Kravtsov}, A.~V., {Klypin}, A.~A., {Primack}, J.~R., \& {Dekel}, A. 2001,
  \mnras, 321, 559

\bibitem[{{Capak} {et~al.}(2004){Capak}, {Cowie}, {Hu}, {Barger}, {Dickinson},
  {Fernandez}, {Giavalisco}, {Komiyama}, {Kretchmer}, {McNally}, {Miyazaki},
  {Okamura}, \& {Stern}}]{2004AJ....127..180C}
{Capak}, P. {et~al.} 2004, \aj, 127, 180

\bibitem[{{Carlberg} {et~al.}(2001){Carlberg}, {Yee}, {Morris}, {Lin}, {Hall},
  {Patton}, {Sawicki}, \& {Shepherd}}]{2001ApJ...552..427C}
{Carlberg}, R.~G., {Yee}, H.~K.~C., {Morris}, S.~L., {Lin}, H., {Hall}, P.~B.,
  {Patton}, D.~R., {Sawicki}, M., \& {Shepherd}, C.~W. 2001, \apj, 552, 427

\bibitem[{{Clowe} {et~al.}(2004){Clowe}, {Gonzalez}, \&
  {Markevitch}}]{2004ApJ...604..596C}
{Clowe}, D., {Gonzalez}, A., \& {Markevitch}, M. 2004, \apj, 604, 596

\bibitem[{{Clowe} \& {Schneider}(2001)}]{2001A&A...379..384C}
{Clowe}, D., \& {Schneider}, P. 2001, \aap, 379, 384

\bibitem[{{Coe} {et~al.}(2006){Coe}, {Ben{\'{\i}}tez}, {S{\'a}nchez}, {Jee},
  {Bouwens}, \& {Ford}}]{2006AJ....132..926C}
{Coe}, D., {Ben{\'{\i}}tez}, N., {S{\'a}nchez}, S.~F., {Jee}, M., {Bouwens},
  R., \& {Ford}, H. 2006, \aj, 132, 926

\bibitem[{{Diaferio} {et~al.}(2005){Diaferio}, {Geller}, \&
  {Rines}}]{2005ApJ...628L..97D}
{Diaferio}, A., {Geller}, M.~J., \& {Rines}, K.~J. 2005, \apjl, 628, L97

\bibitem[{{Diego} {et~al.}(2005){Diego}, {Sandvik}, {Protopapas}, {Tegmark},
  {Ben{\'{\i}}tez}, \& {Broadhurst}}]{2005MNRAS.362.1247D}
{Diego}, J.~M., {Sandvik}, H.~B., {Protopapas}, P., {Tegmark}, M.,
  {Ben{\'{\i}}tez}, N., \& {Broadhurst}, T. 2005, \mnras, 362, 1247

\bibitem[{{Erben} {et~al.}(2001){Erben}, {Van Waerbeke}, {Bertin}, {Mellier},
  \& {Schneider}}]{2001A&A...366..717E}
{Erben}, T., {Van Waerbeke}, L., {Bertin}, E., {Mellier}, Y., \& {Schneider},
  P. 2001, \aap, 366, 717

\bibitem[{{Frye} {et~al.}(2002){Frye}, {Broadhurst}, \&
  {Ben{\'{\i}}tez}}]{2002ApJ...568..558F}
{Frye}, B., {Broadhurst}, T., \& {Ben{\'{\i}}tez}, N. 2002, \apj, 568, 558

\bibitem[{{Fukugita} {et~al.}(1990){Fukugita}, {Futamase}, \&
  {Kasai}}]{1990MNRAS.246P..24F}
{Fukugita}, M., {Futamase}, T., \& {Kasai}, M. 1990, \mnras, 246, 24P

\bibitem[{{Gaidos}(1997)}]{1997AJ....113..117G}
{Gaidos}, E.~J. 1997, \aj, 113, 117

\bibitem[{{Garilli} {et~al.}(1999){Garilli}, {Maccagni}, \&
  {Andreon}}]{1999A&A...342..408G}
{Garilli}, B., {Maccagni}, D., \& {Andreon}, S. 1999, \aap, 342, 408

\bibitem[{{Gavazzi} {et~al.}(2003){Gavazzi}, {Fort}, {Mellier}, {Pell{\'o}}, \&
  {Dantel-Fort}}]{2003A&A...403...11G}
{Gavazzi}, R., {Fort}, B., {Mellier}, Y., {Pell{\'o}}, R., \& {Dantel-Fort}, M.
  2003, \aap, 403, 11

\bibitem[{{Halkola} {et~al.}(2006){Halkola}, {Seitz}, \&
  {Pannella}}]{2006astro.ph..5470H}
{Halkola}, A., {Seitz}, S., \& {Pannella}, M. 2006, ArXiv Astrophysics
e-prints, arXiv:astro-ph/0605470

\bibitem[{{Hamana} {et~al.}(2003){Hamana}, {Miyazaki}, {Shimasaku}, {Furusawa},
  {Doi}, {Hamabe}, {Imi}, {Kimura}, {Komiyama}, {Nakata}, {Okada}, {Okamura},
  {Ouchi}, {Sekiguchi}, {Yagi}, \& {Yasuda}}]{2003ApJ...597...98H}
{Hamana}, T. {et~al.} 2003, \apj, 597, 98

\bibitem[{{Hammer} {et~al.}(1997){Hammer}, {Gioia}, {Shaya}, {Teyssandier}, {Le
  Fevre}, \& {Luppino}}]{1997ApJ...491..477H}
{Hammer}, F., {Gioia}, I.~M., {Shaya}, E.~J., {Teyssandier}, P., {Le Fevre},
  O., \& {Luppino}, G.~A. 1997, \apj, 491, 477

\bibitem[{{Hansen} {et~al.}(2005){Hansen}, {McKay}, {Wechsler}, {Annis},
  {Sheldon}, \& {Kimball}}]{2005ApJ...633..122H}
{Hansen}, S.~M., {McKay}, T.~A., {Wechsler}, R.~H., {Annis}, J., {Sheldon},
  E.~S., \& {Kimball}, A. 2005, \apj, 633, 122

\bibitem[{{Hennawi} {et~al.}(2005){Hennawi}, {Dalal}, {Bode}, \&
  {Ostriker}}]{2005astro.ph..6171H}
{Hennawi}, J.~F., {Dalal}, N., {Bode}, P., \& {Ostriker}, J.~P. 2005, ArXiv
  Astrophysics e-prints, arXiv:astro-ph/0506171

\bibitem[{{Hetterscheidt} {et~al.}(2006){Hetterscheidt}, {Simon}, {Schirmer},
  {Hildebrandt}, {Schrabback}, {Erben}, \& {Schneider}}]{2006astro.ph..6571H}
{Hetterscheidt}, M., {Simon}, P., {Schirmer}, M., {Hildebrandt}, H.,
  {Schrabback}, T., {Erben}, T., \& {Schneider}, P. 2006, arXiv Astrophysics
  e-prints, arXiv:astro-ph/0606571

\bibitem[{{Heymans} {et~al.}(2006){Heymans}, {Van Waerbeke}, {Bacon}, {Berge},
  {Bernstein}, {Bertin}, {Bridle}, {Brown}, {Clowe}, {Dahle}, {Erben}, {Gray},
  {Hetterscheidt}, {Hoekstra}, {Hudelot}, {Jarvis}, {Kuijken}, {Margoniner},
  {Massey}, {Mellier}, {Nakajima}, {Refregier}, {Rhodes}, {Schrabback}, \&
  {Wittman}}]{2006MNRAS.368.1323H}
{Heymans}, C. {et~al.} 2006, \mnras, 368, 1323

\bibitem[{{Hoekstra} {et~al.}(1998){Hoekstra}, {Franx}, {Kuijken}, \&
  {Squires}}]{1998ApJ...504..636H}
{Hoekstra}, H., {Franx}, M., {Kuijken}, K., \& {Squires}, G. 1998, \apj, 504,
  636

\bibitem[{{Hoekstra} {et~al.}(2006){Hoekstra}, {Mellier}, {van Waerbeke},
  {Semboloni}, {Fu}, {Hudson}, {Parker}, {Tereno}, \&
  {Benabed}}]{2006ApJ...647..116H}
{Hoekstra}, H. {et~al.} 2006, \apj, 647, 116

\bibitem[{{Hudson} {et~al.}(1998){Hudson}, {Gwyn}, {Dahle}, \&
  {Kaiser}}]{1998ApJ...503..531H}
{Hudson}, M.~J., {Gwyn}, S.~D.~J., {Dahle}, H., \& {Kaiser}, N. 1998, \apj,
  503, 531

\bibitem[{{Jee} {et~al.}(2005){Jee}, {White}, {Ben{\'{\i}}tez}, {Ford},
  {Blakeslee}, {Rosati}, {Demarco}, \& {Illingworth}}]{2005ApJ...618...46J}
{Jee}, M.~J., {White}, R.~L., {Ben{\'{\i}}tez}, N., {Ford}, H.~C., {Blakeslee},
  J.~P., {Rosati}, P., {Demarco}, R., \& {Illingworth}, G.~D. 2005, \apj, 618,
  46

\bibitem[{{Kaiser}(1995)}]{1995ApJ...439L...1K}
{Kaiser}, N. 1995, \apjl, 439, L1

\bibitem[{{Kaiser} {et~al.}(1995){Kaiser}, {Squires}, \&
  {Broadhurst}}]{1995ApJ...449..460K}
{Kaiser}, N., {Squires}, G., \& {Broadhurst}, T. 1995, \apj, 449, 460

\bibitem[{{Kauffmann} {et~al.}(1997){Kauffmann}, {Nusser}, \&
  {Steinmetz}}]{1997MNRAS.286..795K}
{Kauffmann}, G., {Nusser}, A., \& {Steinmetz}, M. 1997, \mnras, 286, 795

\bibitem[{{Kneib} {et~al.}(1996){Kneib}, {Ellis}, {Smail}, {Couch}, \&
  {Sharples}}]{1996ApJ...471..643K}
{Kneib}, J.-P., {Ellis}, R.~S., {Smail}, I., {Couch}, W.~J., \& {Sharples},
  R.~M. 1996, \apj, 471, 643

\bibitem[{{Kneib} {et~al.}(2003){Kneib}, {Hudelot}, {Ellis}, {Treu}, {Smith},
  {Marshall}, {Czoske}, {Smail}, \& {Natarajan}}]{2003ApJ...598..804K}
{Kneib}, J.-P. {et~al.} 2003, \apj, 598, 804

\bibitem[{{Markevitch} {et~al.}(2004){Markevitch}, {Gonzalez}, {Clowe},
  {Vikhlinin}, {Forman}, {Jones}, {Murray}, \& {Tucker}}]{2004ApJ...606..819M}
{Markevitch}, M., {Gonzalez}, A.~H., {Clowe}, D., {Vikhlinin}, A., {Forman},
  W., {Jones}, C., {Murray}, S., \& {Tucker}, W. 2004, \apj, 606, 819

\bibitem[{{Markevitch} {et~al.}(2002){Markevitch}, {Gonzalez}, {David},
  {Vikhlinin}, {Murray}, {Forman}, {Jones}, \& {Tucker}}]{2002ApJ...567L..27M}
{Markevitch}, M., {Gonzalez}, A.~H., {David}, L., {Vikhlinin}, A., {Murray},
  S., {Forman}, W., {Jones}, C., \& {Tucker}, W. 2002, \apjl, 567, L27

\bibitem[{{Natarajan} {et~al.}(2002){Natarajan}, {Loeb}, {Kneib}, \&
  {Smail}}]{2002ApJ...580L..17N}
{Natarajan}, P., {Loeb}, A., {Kneib}, J.-P., \& {Smail}, I. 2002, \apjl, 580,
  L17

\bibitem[{{Navarro} {et~al.}(1997){Navarro}, {Frenk}, \&
  {White}}]{1997ApJ...490..493N}
{Navarro}, J.~F., {Frenk}, C.~S., \& {White}, S.~D.~M. 1997, \apj, 490, 493

\bibitem[{{Oguri} {et~al.}(2005){Oguri}, {Takada}, {Umetsu}, \&
  {Broadhurst}}]{2005ApJ...632..841O}
{Oguri}, M., {Takada}, M., {Umetsu}, K., \& {Broadhurst}, T. 2005, \apj, 632,
  841

\bibitem[{{Paolillo} {et~al.}(2001){Paolillo}, {Andreon}, {Longo}, {Puddu},
  {Gal}, {Scaramella}, {Djorgovski}, \& {de Carvalho}}]{2001A&A...367...59P}
{Paolillo}, M., {Andreon}, S., {Longo}, G., {Puddu}, E., {Gal}, R.~R.,
  {Scaramella}, R., {Djorgovski}, S.~G., \& {de Carvalho}, R. 2001, \aap, 367,
  59

\bibitem[{{Popesso} {et~al.}(2005){Popesso}, {B{\"o}hringer}, {Romaniello}, \&
  {Voges}}]{2005A&A...433..415P}
{Popesso}, P., {B{\"o}hringer}, H., {Romaniello}, M., \& {Voges}, W. 2005,
  \aap, 433, 415

\bibitem[{{Pracy} {et~al.}(2005){Pracy}, {Driver}, {De Propris}, {Couch}, \&
  {Nulsen}}]{2005MNRAS.364.1147P}
{Pracy}, M.~B., {Driver}, S.~P., {De Propris}, R., {Couch}, W.~J., \& {Nulsen},
  P.~E.~J. 2005, \mnras, 364, 1147

\bibitem[{{Reiprich} {et~al.}(2004){Reiprich}, {Sarazin}, {Kempner}, \&
  {Tittley}}]{2004ApJ...608..179R}
{Reiprich}, T.~H., {Sarazin}, C.~L., {Kempner}, J.~C., \& {Tittley}, E. 2004,
  \apj, 608, 179

\bibitem[{{Rines} {et~al.}(2003){Rines}, {Geller}, {Kurtz}, \&
  {Diaferio}}]{2003AJ....126.2152R}
{Rines}, K., {Geller}, M.~J., {Kurtz}, M.~J., \& {Diaferio}, A. 2003, \aj, 126,
  2152

\bibitem[{{Sadeh} {et~al.}(2006){Sadeh}, {Rephaeli}, \&
  {Silk}}]{2006MNRAS.368.1583S}
{Sadeh}, S., {Rephaeli}, Y., \& {Silk}, J. 2006, \mnras, 368, 1583

\bibitem[{{Sand} {et~al.}(2002){Sand}, {Treu}, \&
  {Ellis}}]{2002ApJ...574L.129S}
{Sand}, D.~J., {Treu}, T., \& {Ellis}, R.~S. 2002, \apjl, 574, L129

\bibitem[{{Sand} {et~al.}(2004){Sand}, {Treu}, {Smith}, \&
  {Ellis}}]{2004ApJ...604...88S}
{Sand}, D.~J., {Treu}, T., {Smith}, G.~P., \& {Ellis}, R.~S. 2004, \apj, 604,
  88

\bibitem[{{Schechter}(1976)}]{1976ApJ...203..297S}
{Schechter}, P. 1976, \apj, 203, 297

\bibitem[{{Seitz} \& {Schneider}(1997)}]{1997A&A...318..687S}
{Seitz}, C., \& {Schneider}, P. 1997, \aap, 318, 687

\bibitem[{{Sharon} {et~al.}(2005){Sharon}, {Ofek}, {Smith}, {Broadhurst},
  {Maoz}, {Kochanek}, {Oguri}, {Suto}, {Inada}, \&
  {Falco}}]{2005ApJ...629L..73S}
{Sharon}, K. {et~al.} 2005, \apjl, 629, L73

\bibitem[{{Spergel} {et~al.}(2006){Spergel}, {Bean}, {Dore'}, {Nolta},
  {Bennett}, {Hinshaw}, {Jarosik}, {Komatsu}, {Page}, {Peiris}, {Verde},
  {Barnes}, {Halpern}, {Hill}, {Kogut}, {Limon}, {Meyer}, {Odegard}, {Tucker},
  {Weiland}, {Wollack}, \& {Wright}}]{2006astro.ph..3449S}
{Spergel}, D.~N. {et~al.} 2006, ArXiv Astrophysics e-prints, arXiv:astro-ph/0603449

\bibitem[{{Van Waerbeke} {et~al.}(2000){Van Waerbeke}, {Mellier}, {Erben},
  {Cuillandre}, {Bernardeau}, {Maoli}, {Bertin}, {Mc Cracken}, {Le F{\`e}vre},
  {Fort}, {Dantel-Fort}, {Jain}, \& {Schneider}}]{2000A&A...358...30V}
{Van Waerbeke}, L. {et~al.} 2000, \aap, 358, 30

\bibitem[{{Yagi} {et~al.}(2002){Yagi}, {Kashikawa}, {Sekiguchi}, {Doi},
  {Yasuda}, {Shimasaku}, \& {Okamura}}]{2002AJ....123...66Y}
{Yagi}, M., {Kashikawa}, N., {Sekiguchi}, M., {Doi}, M., {Yasuda}, N.,
  {Shimasaku}, K., \& {Okamura}, S. 2002, \aj, 123, 66

\bibitem[{{Zekser} {et~al.}(2006){Zekser}, {White}, {Broadhurst},
  {Ben{\'{\i}}tez}, {Ford}, {Illingworth}, {Blakeslee}, {Postman}, {Jee}, \&
  {Coe}}]{2006ApJ...640..639Z}
{Zekser}, K.~C. {et~al.} 2006, \apj, 640, 639

\end{thebibliography}

\end{document}